\documentclass[12pt,a4paper]{article}
\pdfoutput=1
\usepackage{graphicx}
\usepackage[T1]{fontenc}
\usepackage[sc,osf]{mathpazo}
\usepackage{a4wide}  
\usepackage{latexsym,amsthm,amsfonts,amsmath,mathrsfs,amssymb}
\usepackage[unicode,implicit]{hyperref}
\hypersetup{%
  pdftitle    = {Dyonic multicenter black holes of N=2,d=4 supergravity at arbitrary locations}
  pdfkeywords = {Yang-Mills, gravity, Einstein-Yang-Mills, black hole, monopole, dyon, hair, multicenter, soliton, supergravity, supersymmetry},
  pdfauthor   = {Patrick Meessen, Tomas Ortin, Pedro F. Ramirez},
  plainpages  = true,
  colorlinks  = true,
  citecolor   = blue,
  urlcolor    = red,
  linkcolor   = black
}
\newcommand{\hepth}[1]{{\tt
\href{http://www.arXiv.org/abs/hep-th/#1}{hep-th/#1}}}

\newcommand{\arxiv}[1]{{\tt arXiv:\href{http://www.arXiv.org/abs/#1}{#1}}}
\newcommand{\doi}[1]{{\tt DOI:\href{http://dx.doi.org/#1}{#1}}}
\usepackage{tikz}\newcommand{\FPAUO}[2]{
\tikz[scale=.13,
         Uniovi/.style={color=green!51!blue, fill=green!51!blue}
 ] {
 \fill[Uniovi] (0,0) circle (10);
 \fill[white] (0,7) circle (1.5);
 \draw[Uniovi] (-2,7.5) rectangle (2,5.5);
 \fill[white] (-0.3,6.6) rectangle (0.3,0);   
 \fill[white] ( -0.9,6.2) rectangle (.9 ,5.6);
 \fill[white] (-1.4, 5.2) rectangle (1.4, 4.6);
 \fill[white] (0,0) ellipse (3.5 and 4);
 \fill[Uniovi] (-2.5,0.3) rectangle (2.5,-0.3);
 \fill[Uniovi] (-2,2.3) rectangle (2,1.7);
 \fill[Uniovi] (-2,-2.3) rectangle (2,-1.7);
 \fill[white] (-4.5,5.5) rectangle (-2.7,4.9);
 \fill[white] (-3.9,6.1) rectangle (-3.3,4.3);
 \fill[white] (4.5,5.5) rectangle (2.7,4.9);
 \fill[white] (3.9,6.1) rectangle (3.3,4.3);
 \foreach \x in { 0,..., 3 }
   \foreach \y in { 0,...,\x}
    {
     \fill[white] (-6-\x*0.7+\y*1.4,3.5-\x *1.97) -- (-5.6-\x*0.7+\y*1.4,2.4-\x *1.97) -- (-6.4-\x*0.7+\y*1.4,2.4-\x *1.97) -- cycle;
     \fill[white] (6-\x*0.7+\y*1.4,3.5-\x *1.97) -- (5.6-\x*0.7+\y*1.4,2.4-\x *1.97) -- (6.4-\x*0.7+\y*1.4,2.4-\x *1.97) -- cycle;
   };
 \draw (0,-6) node[
                               text centered, 
                               color=white, 
                               font={\fontsize{8}{4}\sffamily\selectfont}
                             ] {FPAUO-#1/#2};
}} 
\usepackage{pgfplots}    
\makeatletter
\@addtoreset{equation}{section}
\makeatother

\begin{document}

\begin{flushright}
\small
\FPAUO{17}{11}\\
IFT-UAM/CSIC-16-077\\
July 12\textsuperscript{th}, 2017\\
\normalsize
\end{flushright}

\begin{center}

 
{\large {\bf {Dyonic black holes at arbitrary locations}}}
 
\vspace{.5cm}

\renewcommand{\thefootnote}{\alph{footnote}}
{\sl Patrick Meessen$^{2}$}${}^{,}$\footnote{E-mail: {\tt meessenpatrick [at] uniovi.es}},
{\sl Tom\'{a}s Ort\'{\i}n$^{1}$}${}^{,}$\footnote{E-mail: {\tt Tomas.Ortin [at] csic.es}}
{\sl and Pedro F.~Ram\'{\i}rez$^{1}$}${}^{,}$\footnote{E-mail: {\tt p.f.ramirez [at]  csic.es}},

\setcounter{footnote}{0}
\renewcommand{\thefootnote}{\arabic{footnote}}

\vspace{.5cm}

${}^{1}${\it Instituto de F\'{\i}sica Te\'orica UAM/CSIC\\
C/ Nicol\'as Cabrera, 13--15,  C.U.~Cantoblanco, E-28049 Madrid, Spain}\\ 
\vspace{0.3cm}

${}^{2}${\it HEP Theory Group, Departamento de F\'{\i}sica, Universidad de Oviedo\\
  Calle Federico Garc\'{\i}a Lorca, 18, E-33007 Oviedo, Spain}\\

\vspace{.5cm}


{\bf Abstract}

\end{center}

\begin{quotation}
  {\small We construct and study stationary, asymptotically flat multicenter
    solutions describing regular black holes with non-Abelian hair (colored
    magnetic-monopole and dyon fields) in two models of $\mathcal{N}=2,d=4$
    Super-Einstein-Yang-Mills theories: the quadratic model
    $\overline{\mathbb{CP}}^{3}$ and the cubic model ST$[2,6]$, which can be
    embedded in 10-dimensional Heterotic Supergravity. These solutions are
    based on the multicenter dyon recently discovered by one of us, which
    solves the SU$(2)$ Bogomol'nyi and dyon equations on $\mathbb{E}^{3}$.  In
    contrast to the well-known Abelian multicenter solutions, the relative
    positions of the non-Abelian black-hole centers are unconstrained.

    We study necessary conditions on the parameters of the solutions that ensure the regularity of the metric. In the case of the
    $\overline{\mathbb{CP}}^{3}$ model we show that it is enough to require
    the positivity of the ``masses'' of the individual black holes, the
    finiteness of each of their entropies and their superadditivity. In the
    case of the $ST[2,6]$ model we have not been able to show that analogous conditions are sufficient, but we give
    an explicit example of a regular solution describing thousands of non-Abelian
    dyonic black holes in equilibrium at arbitrary relative positions.

    We also construct non-Abelian solutions that interpolate smoothly between
    just two aDS$_{2}\times$S$^{2}$ vacua with different radii
    (\textit{dumbbell solutions}).  
}
\end{quotation}

\newpage
\pagestyle{plain}

\tableofcontents


\section{Introduction}

One of the most fascinating features of extremal black-hole solutions is that
they can be superposed or combined, following certain rules, into solutions
that describe several of these objects in equilibrium. Nowadays, these
solutions are referred to as \textit{multicenter} solutions, to encompass more
general cases in which some of the objects associated to the ``centers'' are
not black holes. 

These solutions exhibit very interesting properties which we are going to
review later on, but the most striking of them is that they exist at all. The
existence of stationary solutions describing several gravitating objects in
equilibrium is commonly (and correctly) attributed to cancellation between
attractive gravitational forces and repulsive electric or magnetic forces.
However, apparently, there are no self-interaction terms for the
electromagnetic fields in the actions of the theories in which these solutions
exist (\textit{e.g.}~in the Einstein-Maxwell theory, which admits the
Majumdar-Papapetrou (MP) solutions \cite{Majumdar:1947eu,kn:P} describing
extremal Reissner-Nordstr\"om black holes in static equilibrium).  It is,
therefore, bewildering that the electromagnetic fields \textit{know} that two
centers with fields that correspond to charges of the same kind must repel
each other.

It is useful to remember what the situation in absence of gravity is like. In
that case, we are used to place point-like charges of arbitrary values at
arbitrary points in space and then find the corresponding electrostatic field
which solves all the Maxwell equations with those sources. This is possible
because the Maxwell field is Abelian and it does not \textit{know} what is the
interaction between those centers nor whether they can be in static
equilibrium or should be hold by other forces in the chosen positions unless
interaction terms such as the worldline actions for charged particles,
embodying the Lorentz force, are added to the theory.

In contrast, in a non-Abelian theory such as General Relativity, the
interaction between two mass centers and their motion is completely determined
by the field equations, as shown by Einstein, Grommer, Infeld, Hoffmann and
others in
Refs.~\cite{Einstein:1927,Einstein:1938yz,Fock:1939,Einstein:1940mt,Einstein:1949,Papapetrou:1951}. General
Relativity \textit{knows} that it is not possible to have two Schwarzschild
black holes in static equilibrium because self-interaction is built-in and
\textit{regular} static multicenter solutions simply do not exist.  

Regularity is, evidently, a very important condition in this discussion
because there are indeed solutions describing an arbitrary number of
Schwarzschild black holes placed at arbitrary points in a straight line: the
Israel-Khan solutions \cite{Israel:1964}. However, these solutions have
conical singularities in the lines that join every two contiguous black-hole
centers, the deficit angle being related to the Newtonian force acting between
them. These singularities can be interpreted as \textit{struts} exerting an
additional force to compensate the gravitational attraction and hold the black
holes in their positions.\footnote{In an infinite periodic array of
  Schwarzschild black holes the total gravitation force over each of them
  vanishes and the conical singularities disappear \cite{Myers:1987}.} Many of
the singularities that occur in multicenter solutions can be interpreted along
the same lines: they show that external forces are needed to hold the
configuration in equilibrium. Therefore, we will be interested in the
conditions required to make the singularities disappear and, ultimately, we
will only consider regular solutions.

As we have stressed, the Maxwell equations in curved backgrounds do not
contain any electromagnetic self-interaction terms. The reason why the MP
solutions are possible must, therefore, lie entirely in the gravitational
interaction and, more specifically, in the electromagnetic interaction energy
which is implicitly contained in the electromagnetic energy-momentum
tensor. Gravity may not know directly about electromagnetic interactions
between charged particles but it does know about all the interaction
energies. In the end, this is equivalent to knowing the interactions
themselves well enough as to determine the equations of motion of the mass
centers, as shown by Einstein \textit{et alia}, and also of \textit{charge
  centers}, as shown by Wallace and Infeld in the interesting but less well
known Refs.~\cite{Wallace:1940,Infeld:1940,Wallace:1941}.\footnote{In
  Ref.~\cite{Brill:1963}, Brill and Lindquist studied the time-symmetric
  initial-date problem for several non-extremal Reissner-Nordstr\"om black
  holes and considered the contribution to the total energy in the common
  asymptotically-flat region of the gravitational and electrostatic interaction
  energies, but no connection between their values and the possibility of
  evolving the initial data into a completely regular static solution (a MP
  solution) was made. Similar solutions for the time-symmetric initial-data
  problem in Einstein-Maxwell-dilaton gravity and in models of
  $\mathcal{N}=2,d=4$ supergravity are known \cite{Ortin:1995,Cvetic:2014} and
  could also be studied from the same point of view.}

This mechanism is, obviously, much more general and explains, for instance,
the existence of static multi-D-brane solutions in superstring theory
effective field theories (supergravities) in spite of the fact that, in the
underlying fundamental theory, D-branes have very complex interactions with a
very delicate cancellation associated to supersymmetry
\cite{Polchinski:1995mt}.

The existence of static multicenter solutions, that we have almost grown used
to take for granted, is, therefore, a small wonder and a signal that
supergravities and other effective field theories including gravity encode a
large amount of information of the original (superstring or other) theory.

The equilibrium of forces (or, actually, of interaction energies) required by
these solutions can be achieved in more complicated situations, giving rise to
stationary multicenter solutions. The first family of solutions of this kind
was found by Perj\'es, Israel and Wilson (PIW)
\cite{Perjes:1971gv,Israel:1972vx} in the Einstein-Maxwell theory. They
are an extension of the MP family in which each center can have higher momenta
of the gravitational and electromagnetic fields and the whole spacetime can
also global momenta. If one wants the centers to be regular black
holes\footnote{In the Einstein-Maxwell theory this seems to be the only way to
  avoid having naked singularities.} the dipole and higher momenta of each
center must vanish, in agreement with the no-hair theorem.\footnote{This is
  also the only way to have globally defined unbroken supersymmetry
  \cite{Bellorin:2006xr}.}  Still, these solutions can present other
pathologies such as Misner strings or closed timelike curves (CTCs) and Hartle
and Hawking proved in Ref.~\cite{Hartle:1972ya} that the only regular
solutions in the PIW family are those of the MP subfamily.\footnote{We review
  this result in Appendix~\ref{sec:MinSugr} from our own point of view.}

Generalizations of the PIW family were found in pure $\mathcal{N}=4,d=4$
supergravity (``SWIP'' \cite{Tod:1995jf,Bergshoeff:1996gg,Bellorin:2005zc})
and later in $\mathcal{N}=2,d=4$ supergravity coupled to vector multiplets,
\cite{Behrndt:1997ny,LopesCardoso:2000qm,Meessen:2006tu} typically as timelike
supersymmetric solutions. However, these families were only used to construct
static (MP-like) multicenter solutions and the full potential of these
solutions remained hidden until Denef and Bates showed in
Refs.~\cite{Denef:2000nb,Bates:2003vx} how to construct completely regular
multicenter solutions describing many static black holes with electric and
magnetic charges (dyons) with global angular momentum in $\mathcal{N}=2,d=4$
supergravity coupled to vector multiplets.  

The source for the angular momentum of these solutions is the angular momentum
of the electromagnetic fields due to the presence of Dirac monopoles and
electric charges in different places. As a matter of fact, the global angular
momentum is proportional to the symplectic-invariant Dirac-Schwinger-Zwanziger
quantization condition for dyons and it should, therefore, be
quantized. Generically, these solutions have Misner strings (the gravitational
analog of Dirac strings) which can only be avoided at the price of introducing
CTCs \cite{Misner:1963fr}. The cancellation of the sources of Misner strings
imposes constraints on the charges and location of the centers\footnote{We
  will review these constraints in Section~\ref{sec-setup}.}. These
constraints are extremely hard to solve for 3 or more centers.

The fact that all the multicenter solutions mentioned so far carry Abelian
dyonic charges only suggests that a possible reason for the typical presence
of Misner strings is, precisely, the Abelian nature of the fields and it also
suggests that they could be avoided by the use of non-Abelian fields. However,
no non-Abelian multicenter families of solutions have been constructed so far
and it is the purpose of this paper to do it for the first
time.\footnote{\label{foot-1} A 2-center solution that describes two SU$(2)$
  gravitating BPS magnetic monopoles in equilibrium (an 't~Hooft-Polyakov and a
  Wu-Yang monopole) in $\mathcal{N}=2,d=4$ Super-Einstein-Yang-Mills (SEYM)
  theories was constructed in Ref.~\cite{Bueno:2014mea}, using the solutions
  of the Bogomol'nyi equations found in
  Refs.~\cite{Cherkis:2007jm,Cherkis:2007qa}. However, these are just
  particular solutions which are very hard to generalize to a higher number of
  centers. It is also possible to construct solutions with many Wu-Yang
  magnetic monopoles, but these are equivalent, up to a singular SU$(2)$ gauge
  transformation to solutions with as many Dirac monopoles embedded in SU$(2)$
  and should not be considered as genuinely non-Abelian.} Here we are going to
focus on the 4-dimensional case and in a forthcoming paper we will consider
the 5-dimensional one \cite{kn:MOR}, although some of the considerations made
here will also apply to that case.

Finding gravitating solutions with genuinely non-Abelian fields is a very
complicated problem due to the non-linearities of the equations and,
therefore, so far there have been no attempts to construct multicenter
solutions beyond those mentioned in footnote~\ref{foot-1}. Actually, in the
context of the Einstein-Yang-Mills (EYM) and Einstein-Yang-Mills-Higgs (EYMH)
theories, even the single-center solutions are only known numerically
\cite{Volkov:1989fi,Bizon:1990sr}\footnote{ See also See
  Refs.~\cite{Volkov:1998cc,Galtsov:2001myk}.}. This makes them very difficult
to study, interpret and generalize. As argued in Ref.~\cite{Cano:2017sqy},
embedding a field theory in a supersymmetric one provides new tools to tackle
the problem and construct new solutions, specially if one assumes that they
preserve some supersymmetry, but it is necessary to use
$\mathcal{N}>1$ supersymmetry\footnote{That is, more than 4 supercharges.} for
the kind of solutions we are after. In general, the EYM and EYMH theories are
not consistent truncations of any $\mathcal{N}>1$ supergravity, and,
therefore, if we want to use the solution-generating techniques provided by
supersymmetry, we must consider the simplest extended supergravities that
include non-Abelian Yang-Mills fields, which we have called
Super-Einstein-Yang-Mills (SEYM) theories. Typically, they come equipped with
scalar fields that play the r\^ole of (usually adjoint) Higgs fields and,
typically, low values of $\mathcal{N}$ give more freedom to choose the gauge
group. Thus, $\mathcal{N}=2$ is the optimal value in 4 and 5
dimensions.\footnote{The 5-dimensional supergravity theories with 8
  supercharges will be referred to as $\mathcal{N}=1$ theories because it is
  the minimal value in $d=5$.}

$\mathcal{N}=2,d=4$ SEYM theories are theories of $\mathcal{N}=2,d=4$
supergravity coupled to vector multiplets in which some subgroup of the
isometry group of the Special K\"ahler scalar manifold has been gauged. These
theories are the simplest which include YM fields and have a positive semidefinite scalar potential. This forces the timelike supersymmetric solutions
to be asymptotically flat because the asymptotically-DS$_{4}$ ones cannot be
supersymmetric. If the gauge group has an SU$(2)$ factor, it is also possible
to use it to gauge simultaneously the SU$(2)$ factor of the U$(2)$ R-symmetry
group. The resulting theory has a potential that allows for
asymptotically-aDS$_{4}$ solutions, but it is a much more complicated theory
and only a few solutions (none of them describing black holes) are known
\cite{Ortin:2016sih}, even though the most general timelike supersymmetric
solutions have been characterized in Ref.~\cite{Meessen:2012sr}. Further
generalizations are possible in presence of hypermultiplets, but here we are
going to stick to the simplest possibility.\footnote{A short review of these
  theories can be found in Appendix~\ref{app-N2D4SEYM}. More information is
  available in
  Refs.~\cite{Andrianopoli:1996cm,Freedman:2012zz,Ortin:2015hya}.}

The solution-generating methods needed to construct non-Abelian solutions of
$\mathcal{N}=2,d=4$ SEYM theories were found in Ref.~\cite{Hubscher:2008yz}
and they have been successfully applied to construct, in fully analytical
form, several interesting supersymmetric single-center solutions with genuine
non-Abelian hair such as \textit{global monopoles} and extremal static black
holes \cite{Huebscher:2007hj,Meessen:2008kb,Bueno:2014mea,Meessen:2015nla} and
the non-Abelian 2-center solutions mentioned in footnote~\ref{foot-1}. 

As we will see, the supersymmetric solution-generating technique employed
requires solving the non-Abelian Bogomol'nyi equations in
$\mathbb{E}^{3}$ \cite{Bogomolny:1975de}. The solutions to these equations are
BPS magnetic monopoles such as the SU$(2)$ 't~Hooft-Polyakov monopole in the
Bogomol'nyi-Prasad-Sommerfield limit
\cite{'tHooft:1974qc,Polyakov:1974ek,Prasad:1975kr}, the SU$(2)$ Wu-Yang
monopole \cite{Wu:1967vp} or the colored monopoles found by Protogenov as
part of its full classification of the possible spherically-symmetric
solutions \cite{Protogenov:1977tq} and which can be extended to other gauged
groups \cite{Meessen:2015nla}. All of them have been used to construct regular
black holes or gravitating (global) monopoles in $\mathcal{N}=2,d=4$ SEYM
theories. The multicenter solutions of the Bogomol'nyi equations are expected
to describe several of these magnetic monopoles in equilibrium but only the
very restricted or trivial examples discussed in footnote~\ref{foot-1} were
known until very recently.

In Ref.~\cite{Ramirez:2016tqc}, based on the results of
Ref.~\cite{Etesi:2002cc}, one of us found a multicenter solution of the
SU$(2)$ Bogomol'nyi equations on $\mathbb{E}^{3}$ describing an arbitrary
number of colored monopoles in equilibrium. Furthermore, this configuration was generalized to describe colored dyons through the inclusion of electric non-Abelian sources. This \textit{multi-colored dyon} solution will
provide the basis to construct non-Abelian multicenter solutions in
$\mathcal{N}=2,d=4$ SEYM theories.

Colored magnetic monopoles are very interesting solutions that behave as
Wu-Yang monopoles near the origin but have asymptotically vanishing magnetic
monopole charge.\footnote{This behavior is the source of some interesting
  puzzles involving non-Abelian hair and the entropy of the black holes. The
  solution to this puzzle in the $d=5$ case has been found in
  Ref.~\cite{Cano:2017qrq} in the context of string theory and we are
  currently working on the $d=4$ case \cite{kn:CMOR}. We will not discuss it
  any further here.} In Ref.~\cite{Bueno:2015wva} we showed that they are
related via dimensional oxidation \textit{\`a la Kronheimer}
\cite{kn:KronheimerMScThesis} to the BPST instanton \cite{Belavin:1975fg} and
the multi-colored monopole solution corresponds to a multi-instanton solution in a non-trivial hyper-K\"ahler space \cite{kn:MOR, Etesi:2002cc}.

As in the Abelian case, multi-colored dyon solutions are stationary, rather than static. However, as we are going to see, these never
gives rise to Misner strings and the positions of the dyons can be chosen
completely at will. This is one of the main properties of the non-Abelian
multicenter solutions that we are going to construct here. Another important
property is that, due to the rapid fall-off of the non-Abelian fields at
spatial infinity, the non-Abelian field do not give a net contribution to
the global angular momentum.

The regularity of multicenter solutions is not guaranteed by the absence of
Misner strings alone. It is necessary to study the complete metric and, in
particular, the so-called ``metric function'' $e^{-2U}$ defined in
Eq.~(\ref{eq:conformastationary}), whose behavior determines the regularity
of the black-hole horizon at each center and which must not vanish anywhere
else. We are going to look for general conditions guaranteeing that this is
the case and, at least for some models, we are going to see that they have
very reasonable physical interpretations.

This paper is organized as follows: in Section~\ref{sec-setup} we set up the
problem of finding non-Abelian, timelike supersymmetric, multicenter solutions
of $\mathcal{N}=2,d=4,5$ SEYM theories, introducing the multi-colored dyon
solution. In Section~\ref{sec-n2d4seymsolutions} we focus on the 4-dimensional
case and apply the technique to two models of SU$(2)$ $\mathcal{N}=2,d=4$ SEYM
(the $\overline{\mathbb{CP}}^{3}$ model in Section~\ref{sec-CP3} and the
$\mathrm{ST}[2,6]$ model, which can be embedded in Heterotic Supergravity, in
Section~\ref{sec-st26}), finding solutions whose regularity conditions we will
study in full detail in terms of masses and
entropies. Section~\ref{sec-conclusions} contains our conclusions. In
Appendix~\ref{app-N2D4SEYM} we briefly review $\mathcal{N}=2$, $d=4$ SEYM
theories. In Appendix~\ref{sec:MinSugr} we revise Hartle and Hawking's result
on the non-existence of stationary multi-black-hole solutions in the
Einstein-Maxwell theory (which is just the bosonic sector of pure
$\mathcal{N}=2,d=4$ supergravity).

\section{Setting up the problem}
\label{sec-setup}

The problem of finding timelike supersymmetric solutions of
$\mathcal{N}=2,d=4$ SEYM theories and timelike or null supersymmetric
solutions with an additional isometry of $\mathcal{N}=1,d=5$ SEYM
theories\footnote{These theories are briefly reviewed in
  Appendix~\ref{app-N2D4SEYM}. The $\mathcal{N}=1,d=5$ SEYM case will be dealt
  with in a forthcoming paper \cite{kn:MOR}.} boils down to the far simpler
problem of finding functions $\Phi^{\Lambda},\Phi_{\Lambda}$ and vector fields
$\breve{A}^{\Lambda}{}_{\underline{r}}$\footnote{$\Lambda,\Sigma,\ldots=0,1,\cdots,n_{V4}$
  where $n_{V4}$ is the number of vector supermultiplets in $d=4$ and
  $r,s,\ldots=1,2,3$.} in Euclidean 3-dimensional space $\mathbb{E}^{3}$
solving the following three sets of equations:

\begin{eqnarray}
\label{eq:B}
\tfrac{1}{2}\varepsilon_{\underline{r}\underline{s}\underline{w}}
\breve{F}^{\Lambda}{}_{\underline{s}\underline{w}}
-\breve{\mathfrak{D}}_{\underline{r}}\Phi^{\Lambda}
& = &
0\, ,
\\
& & \nonumber \\
\label{eq:D2Phi}
\breve{\mathfrak{D}}_{\underline{r}}\breve{\mathfrak{D}}_{\underline{r}}\Phi_{\Lambda} 
-g^{2} f_{\Lambda\Sigma}{}^{\Omega}f_{\Delta\Omega}{}^{\Gamma}
\Phi^{\Sigma}\Phi^{\Delta}\Phi_{\Gamma}
& = &
0\, ,
\\
& & \nonumber \\
\label{eq:integrability}
\Phi_{\Lambda}\breve{\mathfrak{D}}_{\underline{r}}\breve{\mathfrak{D}}_{\underline{r}}\Phi^{\Lambda}
-
\Phi^{\Lambda}\breve{\mathfrak{D}}_{\underline{r}}\breve{\mathfrak{D}}_{\underline{r}}\Phi_{\Lambda}
& = &
0\, ,
\end{eqnarray}

\noindent
where $\breve{\mathfrak{D}}_{\underline{r}}$ is the gauge covariant derivative
in $\mathbb{E}^{3}$ with respect to the connection
$\breve{A}^{\Lambda}{}_{\underline{r}}$.

The first set of equations (\ref{eq:B}) are just the Bogomol'nyi equations
\cite{Bogomolny:1975de} for a set of real, adjoint, Higgs fields
$\Phi^{\Lambda}$ and gauge vector fields
$\breve{A}^{\Lambda}{}_{\underline{r}}$ on $\mathbb{E}^{3}$. Due to their
non-linear structure (when the gauge group is non-Abelian) one has to solve
simultaneously for $\Phi^{\Lambda}$ and
$\breve{A}^{\Lambda}{}_{\underline{r}}$. In the Abelian case, the
integrability condition for these equations is the Laplace equation in
$\mathbb{E}^{3}$, \emph{i.e.}
$\partial_{\underline{r}}\partial_{\underline{r}}\Phi^{\Lambda}=0$; the Abelian vector
fields are completely determined by the choice of harmonic functions
$\Phi^{\Lambda}$ and usually they are not written down explicitly.

For the $\mathrm{SU}(2)$ gauge group, which will be our main interest, all the
spherically-symmetric solutions were found by Protogenov in
Ref.~\cite{Protogenov:1977tq}. The BPS limit of the 't~Hooft-Polyakov monopole
\cite{'tHooft:1974qc,Polyakov:1974ek,Prasad:1975kr}, the $\mathrm{SU}(2)$
Wu-Yang monopole \cite{Wu:1967vp} and the so-called \textit{colored monopoles}
considered in Refs.~\cite{Meessen:2008kb,Meessen:2015nla} are, perhaps, the
most interesting solutions. Only the 't~Hooft-Polyakov monopole is regular,
but, just as in the Abelian case, the singularity of the solution in
$\mathbb{E}^{3}$ needs not imply the existence of a spacetime singularity in
the complete supergravity solutions. Actually, the singularities are typically
associated to extremal black hole horizons.

Multicenter solutions of these equations, specially with the right properties
necessary to construct multi-black-hole solutions, are extremely hard to find.
In this paper we will use the multicenter solutions of the Bogomol'nyi
equations found by one of us in Ref.~\cite{Ramirez:2016tqc} to construct multi-center
black-hole solutions in 4 and 5 dimensions. This solution, which will be reviewed in the next section, is based on the multi-instanton
solutions of Etesi and Hausel Ref.~\cite{Etesi:2002cc} and on the general relation between
instantons in hyperK\"ahler spaces and BPS monopoles on $\mathbb{E}^{3}$ found
by Kronheimer in Ref.~\cite{kn:KronheimerMScThesis}.

The second set of equations (\ref{eq:D2Phi}) is a set of linear equations for
the scalar fields $\Phi_{\Lambda}$. For $\mathrm{SU}(2)$ and, more generally,
for compact groups, one can always use the trivial solution $\Phi_{\Lambda} \propto \Phi^{\Lambda}$, which also satisfy
Eqs.~(\ref{eq:integrability}). However, a more interesting set of solutions
has been found in Ref.~\cite{Ramirez:2016tqc} and we will make use of them. In
the Abelian case, again, the $\Phi_{\Lambda}$ are harmonic functions in
$\mathbb{E}^{3}$:
$\partial_{\underline{r}}\partial_{\underline{r}}\Phi_{\Lambda}=0$.

The third equation, (\ref{eq:integrability}) is the integrability condition of
the equations that defines the 1-form $\omega_{\underline{r}}$ that appears in
the 4- and 5-dimensional metrics.  If we use the other two sets of equations,
it seems to be automatically satisfied. However, since, typically, the fields
$\Phi^{\Lambda},\Phi_{\Lambda}$ have singularities, the first two sets of
equations may not be identically satisfied at the locus of the
singularities. When this happens, the 1-form $\omega_{\underline{r}}$ still
exists, but it can only be defined locally: it will exhibit Dirac-Misner
string singularities \cite{Misner:1963fr} that can only be cured by defining
different $\omega_{\underline{r}}$ which are regular in different patches and
identifying these solutions in the overlaps up to ``gauge transformations''
that can be identified as coordinate transformations in the time
direction. The consistency of these construction requires a periodic
identification of the time coordinate with the consequent loss of asymptotic
flatness. For this reason, Eq.~(\ref{eq:integrability}) is required to hold
everywhere and, at the loci of the singularities, this condition leads to
non-trivial equations in the Abelian case which generically
(for non-trivial $\Phi_{\Lambda}$) constrain the relative distances of the
pairs of black holes in terms of their charges and the moduli
\cite{Denef:2000nb,Bates:2003vx}. We will see that the solutions found in
Ref.~\cite{Ramirez:2016tqc} do not imply any such constraints because they
solve identically Eqs.~(\ref{eq:B}-\ref{eq:integrability}) at the would-be singularities.

Given a solution
$\Phi^{\Lambda},\Phi_{\Lambda},\breve{A}^{\Lambda}{}_{\underline{r}}$ of the
above equations there are three sets of rules that allow us to construct timelike supersymmetric solution of $\mathcal{N}=2,d=4$ SEYM
theories and timelike or null solutions with an additional isometry of
$\mathcal{N}=1,d=5$ SEYM theories respectively. The functions and 1-forms
$\Phi^{\Lambda},\Phi_{\Lambda},\breve{A}^{\Lambda}{}_{\underline{r}}$ will be
the building blocks of the physical fields of the solutions. We will review
the rules for the 4-dimensional case in Section~\ref{sec-n2d4seymsolutions}
where we will construct and study explicit solutions of several supergravity
models with a single non-Abelian SU$(2)$ sector. Now we are going to set up
the general problem of solving those equations and we are going to review the
solutions found in Ref.~\cite{Ramirez:2016tqc} to which we will henceforth
refer to as \textit{the multi-colored dyon}.

\subsection{The multi-colored dyon solution}
\label{sec-multicolouredmonopolesolutions}

The indices $\Lambda,\Sigma,\ldots$ that label the vector fields can be split
into those corresponding to the Abelian and non-Abelian ($\mathrm{SU}(2)$)
sectors. Labeling the former with $\lambda,\sigma,\ldots$ and the latter with $A,B,\ldots$, which
will only take three values\footnote{\label{footnote:aindices} We can always
  call these values $1,2,3$ for convenience. Then, we can use the same labels
  for the Cartesian coordinates in $\mathbb{E}^{3}$, which simplifies
  considerably the notation.}. The equations
(\ref{eq:B})-(\ref{eq:integrability}) become\footnote{Our conventions for the
  SU$(2)$ objects are as follows: the structure constants are
  $f_{AB}{}^{C}=+\varepsilon_{ABC}=+\varepsilon_{AB}{}^{C}$ (the upper or
  lower position of the indices, which we will choose for essentially
  esthetic reasons, is irrelevant) and the covariant derivative and gauge
  field strength are
\begin{equation}
\breve{\mathfrak{D}}_{\underline{m}}\Phi^{A}
 = 
\partial_{\underline{m}}\Phi^{A}
+g\varepsilon^{A}{}_{BC}\breve{A}^{B}{}_{\underline{m}}\Phi^{C}\, ,
\qquad
\breve{F}^A\,_{\underline{m}\underline{n}}
 = 
2\partial_{[\underline{m}}\breve{A}^{A}{}_{\underline{n}]}
+g\varepsilon^{A}{}_{BC}\breve{A}^{B}{}_{\underline{m}}
\breve{A}^{C}{}_{\underline{n}}\, .    
\end{equation}
In some cases we use the following vector notation
\begin{equation}
\breve{\mathfrak{D}}_{\underline{m}}\vec{\Phi}
 =  
\partial_{\underline{m}}\vec{\Phi}
+g\breve{\vec{A}}_{\underline{m}}\times \vec{\Phi}\, ,
\qquad
\breve{\vec{F}}_{\underline{m}\underline{n}}
 = 
2\partial_{[\underline{m}}\breve{\vec{A}}_{\underline{n}]}
+g\breve{\vec{A}}_{\underline{m}}\times
\breve{\vec{A}}_{\underline{n}}\, .    
\end{equation}
We will also use the notation
\begin{equation}
\vec{n}_{i}\equiv\frac{\vec{x}-\vec{x}_{i}}{|\vec{x}-\vec{x}_{i}|}\, .
\hspace{1cm}
\mathcal{J}_{A}
=
-2\Phi_{A}\, .
\end{equation}
}

\begin{eqnarray}
\label{eq:B1}
\tfrac{1}{2}\varepsilon_{\underline{r}\underline{s}\underline{w}}
\breve{F}^{\lambda}{}_{\underline{s}\underline{w}}
-\partial_{\underline{r}}\Phi^{\lambda}
& = &
0\, ,
\\
& & \nonumber \\
\label{eq:B2}
\tfrac{1}{2}\varepsilon_{\underline{r}\underline{s}\underline{w}}
\breve{F}^{A}{}_{\underline{s}\underline{w}}
-\breve{\mathfrak{D}}_{\underline{r}}\Phi^{A}
& = &
0\, ,
\\
& & \nonumber \\
\label{eq:D2Phi1}
\partial_{\underline{r}}\partial_{\underline{r}}\Phi_{\lambda} 
& = &
0\, ,
\\
& & \nonumber \\
\label{eq:D2Phi2}
\breve{\mathfrak{D}}_{\underline{r}}\breve{\mathfrak{D}}_{\underline{r}}\Phi_{A} 
-g^{2} \left(
\Phi^{B}\Phi^{B}\Phi_{A}-\Phi^{A}\Phi^{B}\Phi_{B}
 \right) 
& = &
0\, ,
\\
& & \nonumber \\
\label{eq:integrabilitysplit}
\left(
\Phi_{\lambda}\partial_{\underline{r}}\partial_{\underline{r}}\Phi^{\lambda}
-
\Phi^{\lambda}\partial_{\underline{r}}\partial_{\underline{r}}\Phi_{\lambda}
\right)
+
\left(
\Phi_{A}\breve{\mathfrak{D}}_{\underline{r}}\breve{\mathfrak{D}}_{\underline{r}}\Phi^{A}
-
\Phi^{A}\breve{\mathfrak{D}}_{\underline{r}}\breve{\mathfrak{D}}_{\underline{r}}\Phi_{A}
\right)
& = &
0\, .
\end{eqnarray}

The integrability conditions of Eqs.~(\ref{eq:B1}) are
$\partial_{\underline{r}}\partial_{\underline{r}}\Phi^{\lambda}=0$, which are
solved by harmonic functions in $\mathbb{E}^{3}$, as mentioned above.  The
explicit form of the corresponding Abelian vector fields
$\breve{A}^{\lambda}{}_{\underline{r}}$ will not be required in what
follows. It will be sufficient to know that they
exist. Eqs.~(\ref{eq:D2Phi1}), which are also solved by harmonic functions in
$\mathbb{E}^{3}$, can be interpreted as the integrability conditions of
Abelian Bogomol'nyi equations for dual vector fields $\breve{A}_{\lambda\,
  \underline{r}}$, but we will not need to know their explicit forms, either.

In order to obtain multi-center black-hole solutions, the harmonic functions 
$\Phi^{\lambda},\Phi_{\lambda}$ must be of the form

\begin{equation}
\Phi^{\lambda} = \Phi^{\lambda}_{0} +\sum_{\alpha}\frac{\Phi^{\lambda}{}_{\alpha}}{r_{\alpha}}\, ,
\hspace{1cm}   
\Phi_{\lambda} = \Phi_{\lambda\, 0} +\sum_{\alpha}\frac{\Phi_{\lambda\, \alpha}}{r_{\alpha}}\, ,
\hspace{1cm}   
r_{\alpha}\equiv |\vec{x}-\vec{x}_{\alpha}|\, ,
\end{equation}

\noindent
for some points $\vec{x}_{\alpha}$ whose positions may be constrained by the integrability
equations (\ref{eq:integrabilitysplit}).\footnote{There are a number of
  reasons why this is the only possible choice if one wants to construct
  regular 4-dimensional multi-center black-hole solutions. See
  \textit{e.g.}~\cite{Bellorin:2006xr}.}

As shown in Ref.~\cite{Ramirez:2016tqc}, Eqs.~(\ref{eq:B2}) are solved
by\footnote{We will write, from now on $\Phi^{A} =
  -\frac{1}{gP}\partial_{A}P$. See footnote~\ref{footnote:aindices}.}

\begin{equation}
\Phi^{A} 
=
-\delta^{A\underline{r}}\frac{1}{gP}\partial_{\underline{r}}P\, ,
\hspace{1cm}
\breve{A}^{A}{}_{\underline{r}}
=
-\varepsilon^{A}{}_{rs}\frac{1}{gP}\partial_{\underline{s}}P\, ,  
\end{equation}

\noindent
for real functions $P$ satisfying

\begin{equation}
\frac{1}{P}\partial_{\underline{r}}\partial_{\underline{r}}P =0\, . 
\end{equation}

Harmonic functions $P$ of the form

\begin{equation}
\label{eq:multipoleP}
P=P_{0}+\sum_{\alpha}\frac{P_{\alpha}}{r_{\alpha}}\, ,  
\end{equation}

\noindent
satisfy the above equation everywhere in $\mathbb{E}^{3}$, including at the
locus of the singularities $\vec{x}=\vec{x}_{\alpha}$. For just one singularity
($\vec{x}_{1}=0$) and positive coefficients $P_{0},P_{1}$, the corresponding
solution of the Bogomol'nyi equations

\begin{equation}
\Phi^{A} 
=
\frac{1}{gr(1+\lambda^{2}r)} \frac{x^{A}}{r}\, ,
\hspace{1cm}
\breve{A}^{A}{}_{B}
=
\varepsilon^{A}{}_{BC}\frac{1}{gr(1+\lambda^{2}r)} \frac{x^{C}}{r}\, ,  
\hspace{1cm}
\lambda^{2}= P_{0}/P_{1}\, ,  
\end{equation}

\noindent
corresponds to a \textit{colored monopole}
\cite{Meessen:2008kb,Meessen:2015nla}. The behavior of the gauge fields at
infinity is such that using the standard definition of magnetic charge one
gets zero. The non-Abelian fields, in fact, do not seem to contribute to any of the
conserved charges defined at spatial infinity (mass or angular
momentum, as we are going to see). The behavior of the gauge fields near the
singularity $r=0$, though, is the same as in the $\mathrm{SU}(2)$ Wu-Yang
monopole case and they seem to contribute to the quantities that can be defined in the
near-horizon limit, such as the Bekenstein-Hawking entropy, in exactly the
same way as the Abelian fields corresponding to electric or magnetic
charges. Then, one would naively conclude that the addition of a colored monopole to an Abelian
black hole does not modify the asymptotic behavior (a clear violation of the
no-hair and uniqueness ``theorems'') but it does modify the entropy,
diminishing it both in 4 and in 5 dimensions \cite{Bueno:2014mea,Meessen:2015enl,Ortin:2016bnl}. However we have recently shown that, at least in the simpler 5-dimensional cases studied in \cite{Cano:2017sqy, Cano:2017qrq}, this is just an illusion caused by an inadequate identification of the charges of the solution in terms of fundamental objects in string theory; actually the non-Abelian sources modify the asymptotic charges but not the entropy. We expect this to be the appropriate interpretation in more complex configurations as well \cite{kn:CMOR}.

Let us now consider Eqs.~(\ref{eq:D2Phi2}). Apart from the
trivial possibility $\Phi_{a}=K\Phi^{a}$, the following solutions were found in
Ref.~\cite{Ramirez:2016tqc}:

\begin{equation}
\Phi_{A} = -\frac{1}{gP}\partial_{A}Q\, ,
\,\,\,\,\,
\mbox{where}
\,\,\,\,\,
\partial_{A}\left(\frac{1}{P^{2}} \partial_{B}\partial_{B}Q
\right)  
=
0\, .
\end{equation}

\noindent
The simplest way to satisfy this equation is to choose $Q$ as a harmonic
function on $\mathbb{E}^{3}$ with the same poles as $P$:

\begin{equation}
\label{eq:multipoleQ}
Q=Q_{0}+\sum_{\alpha}\frac{Q_{\alpha}}{r_{\alpha}}\, .  
\end{equation}

\noindent
With this choice, Eqs.~(\ref{eq:D2Phi2}) are satisfied everywhere in
$\mathbb{E}^{3}$, including at the singularities of $Q$ and $P$. Since Eqs.~(\ref{eq:B2}),
whose integrability conditions are

\begin{equation}
\breve{\mathfrak{D}}_{\underline{r}}\breve{\mathfrak{D}}_{\underline{r}}\Phi^{A} 
=
0\, ,  
\end{equation}

\noindent
are also satisfied everywhere for the chosen $P$, it is to be expected that
Eq.~(\ref{eq:integrabilitysplit}) do not get any contribution from the
non-Abelian sector. As a matter of fact,

\begin{equation}
\Phi_{A}\breve{\mathfrak{D}}_{\underline{r}}\breve{\mathfrak{D}}_{\underline{r}}\Phi^{A}
-
\Phi^{A}\breve{\mathfrak{D}}_{\underline{r}}\breve{\mathfrak{D}}_{\underline{r}}\Phi_{A}
=
\partial_{\underline{r}}
\left(
\frac{\partial_{\underline{r}}Q \partial_{\underline{s}}\partial_{\underline{s}}P}{P^{2}}
-
\frac{\partial_{\underline{r}}P \partial_{\underline{s}}\partial_{\underline{s}}Q}{P^{2}}
\right)
=
0\, .
\end{equation}

The Abelian sector of Eq.~(\ref{eq:integrabilitysplit}) contains terms proportional to
$\delta^{(3)}(\vec{x}-\vec{x}_{\alpha})$ for all the poles $\vec{x}_{\alpha}$ and we
need them to vanish identically for the reasons explained above. Requiring the
coefficient of each delta function to vanish leads to

\begin{equation}
\Phi_{\lambda\, 0}\Phi^{\lambda}{}_{\alpha}-\Phi^{\lambda}{}_{0}\Phi_{\lambda\, \alpha}
+\sum_{\beta}\frac{\Phi_{\lambda\,
    \beta}\Phi^{\lambda}{}_{\alpha}-\Phi^{\lambda}{}_{M}\Phi_{\lambda\,
    \alpha}}{|\vec{x}_{\beta}-\vec{x}_{\alpha}|}  
=
0\, .
\end{equation}

Summing these equations over the index $N$ we get a constraint relating the
coefficients of the poles $\Phi^{\lambda}{}_{\alpha}$ (which are proportional to
each center's charges) to the constant terms $\Phi^{\lambda}{}_{0}$ which are
related to the values of the scalars at infinity (moduli):

\begin{equation}
\sum_{\alpha}\left(\Phi_{\lambda\, 0}\Phi^{\lambda}{}_{\alpha}-\Phi^{\lambda}{}_{0}\Phi_{\lambda\, \alpha} \right)
=
0\, .  
\end{equation}

\noindent
This condition can be interpreted as requiring the vanishing of the global NUT
charge of the spacetime \cite{Bellorin:2006xr} to avoid global Dirac-Misner
strings or global periodic time. The conditions derived above for each center
have the same meaning and, if the charges have been chosen, they constrain the
relative positions of the centers. These constraints must be compatible with
the triangle inequalities
$|\vec{x}_{\beta}-\vec{x}_{\alpha}|+|\vec{x}_{\alpha}-\vec{x}_{\gamma}| \geq
|\vec{x}_{\beta}-\vec{x}_{\gamma}|$ for any triplet of poles
$\beta,\alpha,\gamma$ and this may not always be possible.  Since our main
interest lies in the non-Abelian sector, we will not discuss these equations in more detail, as they have already been thoroughly studied in the
literature. It suffices to stress that the non-Abelian solution of
Ref.~\cite{Ramirez:2016tqc} does not lead to any restrictions on the relative
positions of the centers whatever the choices of coefficients
$P_{0},P_{\alpha},Q_{0},Q_{\alpha}$.

Since Eqs.~(\ref{eq:integrabilitysplit}) are the integrability conditions of
another set of equations, it is worth taking a look at the solutions of the
latter associated to the choices made here. The equations we are talking about
are those determining the components of the 1-form $\omega_{\underline{r}}$
defined on $\mathbb{E}^{3}$:

\begin{equation}
\label{eq:omega}
\partial_{[\underline{r}}\omega_{\underline{s}]} 
=
2\varepsilon_{rsw}
\left(
\Phi_{\Lambda} \breve{\mathfrak{D}}_{\underline{w}}\Phi^{\Lambda}
-
\Phi^{\Lambda} \breve{\mathfrak{D}}_{\underline{w}}\Phi_{\Lambda}
\right)\, .
\end{equation}

We can write $\omega=\omega^{A}+\omega^{NA}$, where $\omega^{(N)A}$ stands for
the (non)-Abelian contribution:

\begin{eqnarray}
\partial_{[\underline{r}}\omega^{A}_{\underline{s}]} 
& = &
2\varepsilon_{rsw}
\left(
\Phi_{\lambda} \partial_{\underline{w}}\Phi^{\lambda}
-
\Phi^{\lambda} \partial_{\underline{w}}\Phi_{\lambda}
\right)\, ,
\\ 
& & \nonumber \\
\partial_{[\underline{r}}\omega^{NA}_{\underline{s}]} 
& = &
2\varepsilon_{rsw}
\left(
\Phi_{A} \breve{\mathfrak{D}}_{\underline{w}}\Phi^{A}
-
\Phi^{A} \breve{\mathfrak{D}}_{\underline{w}}\Phi_{A}
\right)\, .
\end{eqnarray}

\noindent
If the integrability equations are satisfied, $\omega^{A}$ can be defined in a single
patch. The construction of the exact solutions is reviewed, for instance, in
Ref.~\cite{Ortin:2015hya}. $\omega^{NA}$ was found in
Ref.~\cite{Ramirez:2016tqc} to be given by

\begin{equation}
\label{eq:omegaNA}
\omega^{NA}_{\underline{r}}
=
- 4 \varepsilon_{rsw}\frac{\partial_{\underline{s}}P}{gP} 
\frac{\partial_{\underline{w}}Q}{gP}\, . 
\end{equation}

For $|\vec{x}|>>|\vec{x}_{\alpha}|$, $\omega^{NA}_{\underline{r}}\sim
\mathcal{O}(r^{-5})$; this is too fast to contribute to the asymptotic
charges. Near the center $\vec{x}_{*}$

\begin{equation}
\omega^{NA}_{\underline{r}}
\sim 
-4\varepsilon_{rsw} \frac{(x-x_{*})^{s}}{|\vec{x}-\vec{x}_{*}|} 
\sum_{N\neq *}
\frac{(P_{*}Q_{\alpha}-Q_{*}P_{\alpha})(x_{*}-x_{\alpha})^{w}}{g^2 |x_{*}-x_{\alpha}|^{3}}\, .
\end{equation}

\noindent
In order to determine if $\omega^{NA}$ contributes to the near-horizon limit
we consider  $\omega^{NA}_{\underline{r}}dx^{r}$ in spherical coordinates
centered at $\vec{x}=\vec{x}_{*}$ to find that it
$\omega^{NA}_{\underline{r}}dx^{r}\sim r d\varphi$ where $r$ is the local
radial coordinate. Then, if in this limit the $tt$ component of the
4-dimensional metric  $e^{2U}\sim r^{2}$, the
contributions of $\omega^{NA}$ will be subleading and the solutions will have
the usual aDS$_{2}\times$S$^{2}$ near-horizon limit.

This concludes the general discussion. We are now ready to construct
4-dimensional solutions from the building blocks we have introduced and
studied here.

\section{Solutions of   $\mathcal{N}=2$, $d=4$ SEYM}
\label{sec-n2d4seymsolutions}

Given a solution
$\Phi^{\Lambda},\Phi_{\Lambda},\breve{A}^{\Lambda}{}_{\underline{r}}$ of
Eqs.~(\ref{eq:B})-(\ref{eq:integrability}) a timelike supersymmetric solution
of a $\mathcal{N}=2, d=4$ SEYM theory with $n_{V4}$ vector supermultiplets can
be constructed as follows \cite{Huebscher:2007hj,Hubscher:2008yz}:

\begin{enumerate}

\item The elementary building blocks of the solutions, which are the $2(n_{V4}+1)$
  time-independent functions $(\mathcal{I}^{M}) =\left(
  \begin{smallmatrix}
    \mathcal{I}^{\Lambda} \\ \mathcal{I}_{\Lambda} \\
  \end{smallmatrix}
\right)$
are given by 

\begin{equation}
\mathcal{I}^{\Lambda} = -\sqrt{2}\Phi^{\Lambda}\, ,
\hspace{1cm}
\mathcal{I}_{\Lambda} = -\sqrt{2}\Phi_{\Lambda}\, ,  
\end{equation}

\item Given the functions $\mathcal{I}^{M}$, we must find the 1-form on
  $\mathbb{E}^{3}$ $\omega_{\underline{r}}$ by solving Eq.~(\ref{eq:omega}).

\item To reconstruct the physical fields from the functions $\mathcal{I}^{M}$
  we need to solve the stabilization equations, a.k.a.~\textit{Freudenthal
    duality equations}, which give the components of the Freudenthal
  dual\footnote{In
    Refs.~\cite{Meessen:2006tu,Huebscher:2007hj,Hubscher:2008yz} the
    components of the Freudenthal dual are denoted by $\mathcal{R}^{M}$.}
  $\tilde{\mathcal{I}}^{M}(\mathcal{I})$ in terms of the functions
  $\mathcal{I}^{M}$ \cite{Galli:2012ji}; These relations completely
  characterize the model of $\mathcal{N}=2$, $d=4$ supergravity, but they may
  be not unique \cite{Dominic:2014zia,Manda:2015zoa}.

  Equivalently, the $\tilde{\mathcal{I}}^{M}(\mathcal{I})$ can be derived from
  a homogeneous function of degree 2 called the {\em Hesse
    potential}, $W(\mathcal{I})$, as \cite{Bates:2003vx,Mohaupt:2011aa,Meessen:2011aa}

\begin{equation}
\label{eq:HessePotential}
\tilde{\mathcal{I}}^{M} 
= 
-\tfrac{1}{2}\Omega^{MN}
\frac{\partial W}{\partial \mathcal{I}^{N}} 
\;\;\;\longrightarrow\;\;\;
W(\mathcal{I}) = \Omega_{MN}\mathcal{I}^{M}\tilde{\mathcal{I}}^{N}(\mathcal{I}) \, ,
\end{equation}

\noindent
where $\left(\Omega_{MN} \right) = \left(\Omega^{MN} \right) \equiv
\bigl(\begin{smallmatrix} 0 & \mathbb{I}\\ -\mathbb{I} & 0
\end{smallmatrix} \bigr)$ is the symplectic form.

\item The metric takes the form 

\begin{equation}
\label{eq:conformastationary}
ds^{2}
=
e^{2U} (dt+\omega)^{2} -e^{-2U}dx^{r}dx^{r}\, ,
\end{equation}

\noindent
where $\omega=\omega_{\underline{r}}dx^{r}$ is the above spatial 1-form and
the metric function $e^{-2U}$ is given by the Hesse potential

\begin{equation}
e^{-2U} 
= 
W(\mathcal{I}) \, .
\end{equation}

\item The scalar fields are given by 

\begin{equation}
Z^{i}
=
\frac{\tilde{\mathcal{I}}^{i}+i\mathcal{I}^{i}}{\tilde{\mathcal{I}}^{0}+i\mathcal{I}^{0}}\,
,
\,\,\,\,\,
i=1,\cdots, n_{V4}\, .    
\end{equation}

\item The components of the vector fields are given by 

\begin{eqnarray}
\label{eq:recipe4dvectors1}
A^{\Lambda}{}_{t}
& = &
-\tfrac{1}{\sqrt{2}}e^{2U}\tilde{\mathcal{I}}^{\Lambda}\, ,
\\
& & \nonumber \\
\label{eq:recipe4dvectors2}
A^{\Lambda}{}_{\underline{r}}
& = &
\breve{A}^{\Lambda}{}_{\underline{r}} +\omega_{\underline{r}}\ A^{\Lambda}{}_{t}\, .
\end{eqnarray}

\end{enumerate}

\subsection{Solutions of the $\overline{\mathbb{CP}}^{3}$ model}
\label{sec-CP3}

\subsubsection{The model}

The $\overline{\mathbb{CP}}^{3}$ model is characterized by the quadratic
prepotential

\begin{equation}
\mathcal{F} 
=
-\tfrac{i}{4}\eta_{\Lambda\Sigma}
\mathcal{X}^{\Lambda}\mathcal{X}^{\Sigma},
\hspace{1cm}
(\eta_{\Lambda\Sigma}) = \mathrm{diag}(+---)\, .
\end{equation}

The scalars parametrize the symmetric space U$(1,3)/($U$(1)\times$U$(3))$ and
the whole model is invariant under global $\mathrm{U}(1,3)=
\mathrm{U}(1)\times\mathrm{SU}(1,3)$ transformations. We consider the theory
obtained by gauging the
$\mathrm{SO}(3)\subset\mathrm{SU}(3)\subset\mathrm{SU}(1,3)$
subgroup. $\mathrm{SO}(3)$ acts in the adjoint representation on the three
vector multiplets of the model, that we are going to label with $A,B,\ldots$
so that $\eta_{\Lambda\Sigma} \mathcal{X}^{\Lambda}\mathcal{X}^{\Sigma}=
(\mathcal{X}^{0})^{2}-\mathcal{X}^{A}\mathcal{X}^{A}$.

All we need to construct supersymmetric solutions is the
$\overline{\mathbb{CP}}^{3}$ Hesse potential 

\begin{equation}
\mathsf{W}(\mathcal{I}) 
= 
\tfrac{1}{2}\eta_{\Lambda\Sigma}\mathcal{I}^{\Lambda}\mathcal{I}^{\Sigma}
+2\eta^{\Lambda\Sigma}\mathcal{I}_{\Lambda}\mathcal{I}_{\Sigma}\, .
\end{equation}

More details on these models can be found in
Refs.~\cite{Bueno:2014mea,Ortin:2015hya}.

\subsubsection{The solutions}

The Abelian sector of the model is determined by the complex harmonic function
$\mathcal{H}\equiv\Phi^{0}+2i\Phi_{0}$ and the non-Abelian one by the two
triplets of real functions $\Phi^{A}$ and $\mathcal{J}_{A}\equiv -2\Phi_{A}$.
According to the general discussion, if we use the multi-colored
dyonic solution we only need to solve the Abelian part of the integrability
equations (\ref{eq:integrabilitysplit}). For just one Abelian vector the only possibility is
$\Re\mathfrak{e}\, \mathcal{H}\propto \Im\mathfrak{m}\, \mathcal{H}$ or,
equivalently, $\mathcal{H}=e^{i\gamma}H$ for some real harmonic function $H$
and a constant phase $\gamma$. Then, according to the discussion in Section~\ref{sec-setup}, the solution is given in terms
of three harmonic functions $H,P,Q$ with singularities at the same $N$
isolated points $\vec{x}=\vec{x}_{\alpha}$

\begin{equation}
\label{eq:14}
H
=  
h\ +\ \sum_{\alpha=1}^{N}\ \frac{p_{\alpha}}{r_{\alpha}} \, ,
\hspace{1cm}
P 
=
\lambda+\sum_{\alpha=1}^{N}\frac{s_{\alpha}}{r_{\alpha}}\, ,  
\hspace{1cm}
Q
=
-\sum_{\alpha=1}^{N}\frac{\eta_{\alpha}s_{\alpha}/2}{r_{\alpha}}\, ,
\end{equation}

\noindent
by

\begin{equation}
\label{eq:PhiJ}
\Phi^{0}
= 
-H\, ,
\hspace{1cm}
\vec{\Phi}
=
-\frac{1}{gP}\vec{\nabla}P\, ,
\hspace{1cm}
\vec{\mathcal{J}}
=
\frac{2}{gP}\vec{\nabla}Q\, .
\end{equation}

\noindent
The metric function, the 1-form $\vec{\omega}=(\omega_{\underline{r}})$, the
scalar fields and the scalar potential can be written as


\begin{eqnarray}
\label{eq:15}
e^{-2U} 
& = &  
H^{2}\ -\ \vec{\Phi}^{2}\ -\ \vec{\mathcal{J}}^{2} \, , 
\\
& & \nonumber \\
\label{eq:15c}
\vec{\omega}
& = &
2g^{2}\, \vec{\Phi}\times \vec{\mathcal{J}}\, ,
\\
& & \nonumber \\
\label{eq:15b}
\vec{Z} 
& = & 
e^{-i\gamma} 
\frac{\vec{\Phi} + i\vec{\mathcal{J}}}{H} \, ,
\\
& & \nonumber \\
\label{eq:16}
V 
& = & 
2g^{2} e^{4U}\ |\vec{\Phi}\times\vec{\mathcal{J}}|^{2} \; .
\end{eqnarray}

The vector fields of the solution can be constructed using the general
recipe, Eqs.~(\ref{eq:recipe4dvectors1}) and (\ref{eq:recipe4dvectors2}), but
we will not do it explicitly here as we are more concerned with the regularity
of the metric and scalar fields.

\subsubsection{Spherically-symmetric and dumbbell solutions}

As a warm-up exercise, it is convenient to start by the construction of a
single-center solution of this model, which is static because with a single
center necessarily must have $\vec{\Phi}\propto \vec{\mathcal{J}}$. This was already done in
Ref.~\cite{Meessen:2008kb} (with less independent parameters), but here we
will show that there is also a \textit{Robinson-Bertotti dumbbell solution}
similar to the one recently discovered in a 6-dimensional context in
Ref.~\cite{Cano:2016rls}. These single-center dumbbells are obtained by
setting to zero the constant term in the harmonic functions of the Abelian
sector. Without the non-Abelian colored monopole, we would simply
obtain the standard Robinson-Bertotti aDS$_{2}\times$S$^{2}$ solution, which is sometimes called a double extreme black hole. When the colored dyon is included the geometry gets modified. However, the non-Abelian field decays very fast with the distance and the original aDS$_{2}\times$S$^{2}$ asymptotic is recovered. On the other hand, near the origin, the colored dyon contributes as just another ``Abelian'' charge and one also gets an
aDS$_{2}\times$S$^{2}$ spacetime, albeit with different radius (smaller than the original). Thus,
the Robinson-Bertotti dumbbell solution interpolates between two
aDS$_{2}\times$S$^{2}$ spacetimes of different radii.\footnote{The
  6-dimensional Robinson-Bertotti dumbbell solution found in
  Ref.~\cite{Cano:2016rls} interpolates between two aDS$_{3}\times$S$^{3}$
  spacetimes.}

Taking $N=1$ (and suppressing the indices that label the centers), we get 

\begin{eqnarray}
e^{-2U} 
& = &  
h^{2} +\frac{2hp}{r} 
+\left[p^{2} -\frac{(1+\eta^{2})s^{2}}{g^{2}P^{2}r^{2}} \right]\frac{1}{r^{2}}\, ,
\\
& & \nonumber \\
\vec{Z} 
& = & 
\frac{e^{-i\gamma}(1+i\eta)s}{gPH}\frac{\vec{n}}{r^{2}}\, .
\end{eqnarray}

Let us analyze the asymptotically-flat ($h^{2}=1$) case first. It is convenient to
define

\begin{equation}
M= hp\, ,
\hspace{1cm}
E = p^{2} -(1+\eta^{2})/g^{2}\, ,
\end{equation}

\noindent
in terms of which the metric function takes the form

\begin{equation}
e^{-2U} 
= 
1+ \frac{2M}{r}  
+\left[E +\frac{(1+\eta^{2})}{g^{2}}R(r) \right]\frac{1}{r^{2}}\, ,
\end{equation}

\noindent
where we have defined the manifestly positive function 

\begin{equation}
\label{eq:Rfunction}
R(r) 
\equiv
\frac{(1+\frac{\lambda}{s}r)^{2} 
-1}{(1+\frac{\lambda}{s}r )^{2}}\, ,  
\end{equation}

\noindent
which varies smoothly from $0$ at $r=0$ to $1$ at $r=\infty$.

In the above form the metric function is, therefore, manifestly positive if
$M$ (which is the mass) and $E$ (which will be seen to be the entropy times
$\pi$) are both positive. In the asymptotic and near-horizon limits we find respectively

\begin{equation}
\begin{array}{rclrclrcl}
r & \rightarrow & \infty\, ,\,\,\,\,\,    
&
e^{-2U} & \sim & 1+{\displaystyle\frac{2M}{r}} +\mathcal{O}(r^{-2})\, ,
&
\vec{Z} & \sim & \mathcal{O}(r^{-2})\, ,
\\
& & & & & & & & \\
r & \rightarrow & 0\, ,    
&
e^{-2U} & \sim &  {\displaystyle\frac{E}{r^{2}}}+\mathcal{O}(r^{-1})\, ,
&
\vec{Z} & \sim & {\displaystyle\frac{e^{-i\gamma}(1+i\eta)}{gp}}\vec{n} 
+\mathcal{O}(r^{-1})\, ,
\\
\end{array}
\end{equation}

\noindent
showing that the colored dyon field cannot be seen asymptotically but does
contribute to the near-horizon geometry: \emph{i.e.} it appears in the entropy $E$ and in the
covariant attractor value of the scalars \cite{Huebscher:2007hj}.

Setting $h=0$ (with $\lambda\neq 0$) we get the dumbbell solution

\begin{eqnarray}
e^{-2U} 
& = & 
\left[E +\frac{(1+\eta^{2})}{g^{2}}R(r) \right]\frac{1}{r^{2}}\, ,
\\
& & \nonumber \\
\vec{Z} & \sim & \frac{e^{-i\gamma}(1+i\eta)}{gp(1+\frac{\lambda}{s}r)}\vec{n} \, .
\end{eqnarray}

\noindent
The metric function interpolates smoothly between $E/r^{2}$ at $r\sim 0$ and
$p^{2}/r^{2}$ at $r\sim \infty$ while staying always positive. The scalars
interpolate between two covariantly-constant attractors which have different
$r$-dependence because the gauge connection behaves differently in both
limits.

\subsubsection{Multicenter solutions}

For more than one center the metric function is given by 

\begin{eqnarray}
e^{-2U}
& = & 
h\ +\ \sum_{\alpha=1}^{N}\ \frac{2hp_{\alpha}}{r_{\alpha}} 
\\
& & \nonumber \\
\label{eq:17b}
& & 
+\ \sum_{\alpha=1}^{N}\ 
\left[ 
p_{\alpha}^{2} \ -\ \frac{(1+\eta_{\alpha}^{2})s_{\alpha}^{2} }{g^{2} P^{2}r_{\alpha}^{2} } 
\right]  
\frac{1}{r_{\alpha}^{2}}
\\
& & \nonumber \\
& & 
\label{eq:17c}
+2\sum_{\alpha>\beta}^{N}\ 
\left[
p_{\alpha}p_{\beta}
\ -\ \frac{ (1+\eta_{\alpha}\eta_{\beta})s_{\alpha}s_{\beta}}{g^{2}P^{2}r_{\alpha}r_{\beta}}\ 
\vec{n}_{\alpha}\cdot\vec{n}_{\beta}
\right]\, 
\frac{1}{r_{\alpha}r_{\beta}} \; .
\end{eqnarray}

\noindent
Inspired by the single-center case, we now define

\begin{eqnarray}
\label{eq:18}
M_{\alpha} 
& \equiv & hp_{\alpha} \; , 
\\
& & \nonumber \\
\label{eq:18b}
E_{\alpha} 
& \equiv & 
p_{\alpha}^{2}\ -\ (1+\eta_{\alpha}^{2})/g^{2} \; , 
\\
& & \nonumber \\
\label{eq:18c}
E_{\alpha\beta} 
& \equiv & 
(p_{\alpha}+p_{\beta})^{2}\ -\ 4/g^{2}\ -\ (\eta_{\alpha}+\eta_{\beta})^{2}/g^{2} \; ,
\end{eqnarray}

\noindent
as then the term in line (\ref{eq:17b}) can be expressed as

\begin{equation}
\label{eq:19}
\sum_{\alpha=1}^{N}
\left\{
E_{\alpha}\ +\ \frac{(1+\eta_{\alpha}^{2})}{g^{2}}\ 
\left[1 -\frac{ s_{\alpha}^{2}}{P^{2}r_{\alpha}^{2}} \right]
\right\}
 \; ,
\end{equation}

\noindent
whereas the term in line (\ref{eq:17c}) can be written as

\begin{equation}
  \label{eq:20}
\sum_{\alpha>\beta}^{N}\
\left\{  
E_{\alpha\beta}-E_{\alpha}-E_{\beta}\ +\ 
\frac{2(1+\eta_{\alpha}\eta_{\beta})}{g^{2}}\ 
\left[
1\ -\ \frac{s_{\alpha}s_{\beta}}{P^{2}r_{\alpha}r_{\beta}}\ \vec{n}_{\alpha}\cdot\vec{n}_{\beta}
  \right]
\right\}\frac{1}{r_{\alpha}r_{\beta}} \; .
\end{equation}

The last terms in Eqs.~(\ref{eq:19}) and (\ref{eq:20}) are easily seen to be
positive. First, we define the positive functions $K_{\alpha}$

\begin{equation}
\label{eq:21}
\frac{r_{\alpha}P}{s_{\alpha}} 
\; =\; 
1\ +\ \lambda \frac{r_{\alpha}}{s_{\alpha}}\ 
+\ \sum_{\beta\neq\alpha}\ \frac{r_{\alpha}}{r_{\beta}}\frac{s_{\beta}}{s_{\alpha}}
 \; \equiv\; 
1\ +\ K_{\alpha} \; ,
\end{equation}

\noindent
and then, we write

\begin{eqnarray}
\label{eq:22}
\left[1 - \frac{s_{\alpha}^{2} }{P^{2}r_{\alpha}^{2}} \right] 
& =&
\frac{(1+K_{\alpha})^{2}-1}{(1+K_{\alpha})^{2}}
\equiv 
R_{\alpha}\, ,
\\
& & \nonumber \\
\label{eq:23}
\left[ 
1\ -\ \frac{s_{\alpha}s_{\beta}}{r_{\alpha}r_{\beta}P^{2}}\
\vec{n}_{\alpha}\cdot\vec{n}_{\beta}
\right]
& = &
\frac{(1+K_{\alpha})(1+K_{\beta})-\vec{n}_{\alpha}\cdot\vec{n}_{\beta}}
{(1+K_{\alpha})(1+K_{\beta})}
\equiv 
R_{\alpha\beta}\, ,
\,\,\,\,\,\,\,
R_{\alpha\alpha} = R_{\alpha}\, ,
\end{eqnarray}

\noindent
from which the positivity is paramount because the functions $K_{\gamma}$ are
positive and $\vec{n}_{\alpha}\cdot\vec{n}_{\beta}\in [-1,1]$. Since there is
a term $(1+\eta_{\alpha}\eta_{\beta})$ multiplying the whole second term we
need to impose the condition that 

\begin{equation}
\label{eq:signCP3}
\mathrm{sign}(\eta_{\alpha}) =
\mathrm{sign}(\eta_{\beta}) \, .
\end{equation}

\noindent
The function $R_{\alpha}$ is a generalization
of the function $R$ defined in Eq.~(\ref{eq:Rfunction}) for the single-center
case and varies from $0$ at $r_{\alpha}=0$ to $1$ at infinity or at any other
point $r_{\beta}=0$ $\beta\neq\alpha$. The functions $R_{\alpha\beta}$ are
also bound by $0$ and $1$ and are equal to $1$ at all the points
$r_{\gamma}=0$ and at infinity.

The metric function takes the final form 

\begin{eqnarray}
\label{eq:metricfunctionCPf}
e^{-2U}
& = & 
h\ +\ \sum_{\alpha=1}^{N}\ \frac{2M_{\alpha}}{r_{\alpha}} 
+\ \sum_{\alpha=1}^{N}\ 
\left[ 
E_{\alpha}\ +\ \frac{(1+\eta_{\alpha}^{2})}{g^{2}}\ R_{\alpha}
\right]  
\frac{1}{r_{\alpha}^{2}}
\nonumber \\
& & \nonumber \\
& & 
+\sum_{\alpha>\beta}^{N}\ 
\left[
E_{\alpha\beta}-E_{\alpha}-E_{\beta}\ 
+\ \frac{2(1+\eta_{\alpha}\eta_{\beta})}{g^{2}}\ R_{\alpha\beta}
\right]\, 
\frac{1}{r_{\alpha}r_{\beta}} \; ,
\end{eqnarray}

\noindent
and its positivity can be guaranteed by imposing the conditions for all
$\alpha,\beta$

\begin{equation}
M_{\alpha}>0\, ,
\hspace{1cm}
E_{\alpha}>0\, ,
\hspace{1cm}
E_{\alpha\beta}\geq E_{\alpha}+E_{\beta}\, ,  
\end{equation}

\noindent
and the sign condition \eqref{eq:signCP3}. The only poles in the metrical factor (the zeroes of $g_{tt}=e^{2U}$, and,
hence, the horizons) are the ones at the points $r_{\gamma}=0$. 

As can be seen in the asymptotic expansion $r\rightarrow \infty$, the physical
meaning of the first set of conditions is that the mass that each individual
black hole would have if it were isolated must be positive. The meaning of the
other two sets of conditions comes from the study of the near-horizon limits
$r_{\alpha}\rightarrow 0$. In that limit the dominant term is the coefficient
of $1/r_{\alpha}^{2}$ the value of $\alpha$ we are dealing with. Since
$R_{\alpha}$ vanishes precisely at $r_{\alpha}=0$ only the constant part of
the coefficient, $E_{\alpha}$, survives and we get an aDS$_{2}\times S^{2}$
geometry with metric

\begin{equation}
\label{eq:24}
ds^{2}_{nh}
\; =\; 
\frac{r_{\alpha}^{2}}{E_{\alpha}} dt^{2} \ -\
\frac{E_{\alpha}}{r_{\alpha}^{2}}\ dr_{\alpha}^{2}\ 
-\ E_{\alpha}\ d\Omega_{(2)}^{2} \; ,
\end{equation}

\noindent
so $E_{\alpha}$, as the notation suggests, is the entropy of the $\alpha$'th
black hole up to a factor of $\pi$. Thus, we are asking for all the individual
extremal black holes to have a regular horizon. 

The third set of conditions amounts, then, to the requirement that the entropy
of a black hole whose charges are those of the pair $\alpha\beta$ combined
should be larger than the sum of the individual entropies, \textit{i.e.}~we
are assuming the superadditivity of the entropy.

In some special cases, though, the third set of conditions is more restrictive
than necessary to ensure the regularity of the metric. Notice that the metric function can be positive everywhere even if the second line in Eq.~(\ref{eq:metricfunctionCPf}) has negative constant coefficient. For instance, in the two centers case, the
constant coefficients of the $1/r_{\alpha}^{2}$ and $1/(r_{\alpha}r_{\beta})$
terms are

\begin{equation}
\frac{E_{1}}{r_{1}^{2}}
+\frac{E_{2}}{r_{2}^{2}}
+\left[
E_{12}-E_{1}-E_{2}
\right]\, 
\frac{1}{r_{1}r_{2}}\, ,
\end{equation}

\noindent
and can be rewritten in this form:

\begin{equation}
\left[
\frac{\sqrt{E_{1}}}{r_{1}}
-
\frac{\sqrt{E_{2}}}{r_{2}}
\right]^{2}
+
\left[
E_{12}-\left(\sqrt{E_{1}}-\sqrt{E_{2}}\right)^{2}
\right]\, 
\frac{1}{r_{1}r_{2}}\, ,
\end{equation}

\noindent
This combination is non-negative everywhere if

\begin{equation}
E_{12}\geq \left(\sqrt{E_{1}}-\sqrt{E_{2}}\right)^{2}\, ,
\end{equation}

\noindent
which is a weaker condition for which we have, however, no clear physical
interpretation.

%
%

%

The physical scalars are regular everywhere and can be written as

\begin{equation}
\label{eq:28}
\vec{Z}
\; =\; 
\sum_{\alpha=1}^{N}\
\frac{e^{-i\gamma}(1+i\eta_{\alpha})s_{\alpha}}{gHPr_{\alpha}^{2}}\vec{n}_{\alpha}\, ,
\end{equation}

\noindent
and vanish as $\mathcal{O}(r^{-2})$ at infinity while. At the $\alpha$th
center they take the covariantly-constant attractor value
$\frac{e^{-i\gamma}(1+i\eta_{\alpha})}{gp_{\alpha}}\ \vec{n}_{\alpha}$.

In the previous discussion we have ignored the presence of a non-trivial
1-form $\omega_{\underline{r}}dx^{r}$ in the metric given by
Eqs.~(\ref{eq:omegaNA}) or (\ref{eq:15c}) because, asymptotically, it vanishes
faster than any other function in the metric and, in the near-horizon limits,
they are also subleading. However, we must see if its presence gives rise to
pathologies such as closed timelike curves. For the harmonic functions $P$ and
$Q$ in Eq.~(\ref{eq:14}) it takes the explicit form

\begin{equation}
\vec{\omega} 
= 
-4 \sum_{\alpha>\beta} 
\frac{s_{\alpha}s_{\beta}(\eta_{\beta}-\eta_{\alpha})}{g^2 P^{2}r_{\alpha}^{2}r_{\beta}^{2}} 
\vec{n}_{\alpha}\times \vec{n}_{\beta}
=
-4 \sum_{\alpha>\beta} 
\frac{s_{\alpha}s_{\beta}(\eta_{\beta}-\eta_{\alpha})}{g^2 P^{2}r_{\alpha}^{3}r_{\beta}^{3}} 
[\vec{x}\times \vec{x}_{\alpha\beta} +\vec{x}_{\beta}\times\vec{x}_{\alpha}]\, .
\end{equation}

\noindent
Far away from the centers, and for $\lambda\neq 0$,

\begin{equation}
\vec{\omega} 
\sim
\frac{\vec{v}\times \vec{x}}{r^{6}} \, ,
\,\,\,\,\,
\mbox{where}
\,\,\,\,\,
\vec{v}
\equiv
4 \sum_{\alpha>\beta} 
\frac{s_{\alpha}s_{\beta}(\eta_{\beta}-\eta_{\alpha})\vec{x}_{\alpha\beta}}{g^2 \lambda^{2}}\, ,
\end{equation}

\noindent
and choosing coordinates such that $\vec{v}$ is parallel to the $z$ axis

\begin{equation}
\omega \sim \frac{v(ydx-xdy)}{r^{6}}=
\frac{v\sin^{2}{\theta}d\varphi}{r^{4}}\, , 
\end{equation}

\noindent
and $g_{\varphi\varphi} = e^{2U}\omega_{\varphi}^{2}
-e^{-2U}r^{2}\sin^{2}{\theta}$ is clearly negative in that limit. When
$\lambda=0$, then $\omega_{\varphi}\sim r^{-2}$ asymptotically, decaying still too fast to contribute to the
angular momentum or to modify the sign of $g_{\varphi\varphi}$.

In the near-horizon limit $\vec{x}\rightarrow \vec{x}_{*}$, where $\vec{x}_{*}$ denotes the coordinates of the center we are zooming on

\begin{equation}
\vec{\omega} 
\sim
\frac{\vec{u}_*\times \vec{x}}{r} \, ,
\,\,\,\,\,
\mbox{where}
\,\,\,\,\,
\vec{u}_{*}
\equiv
-\frac{4}{g^2 s_{*}} \sum_{\alpha\neq *} 
\frac{ s_{\alpha}(\eta_{\alpha}-\eta_{*})\vec{x}_{\alpha}}{r_{\alpha *}^{3}}\, .
\end{equation}

\noindent
We can choose adapted coordinates such that now $\vec{u}$ is parallel to the $z$ axis, so we can write to leading order

\begin{equation}
\omega \sim \frac{u(ydx-xdy)}{r}
=
u_{*}r\sin^{2}{\theta}d\varphi \, , 
\end{equation}

\noindent
and 

\begin{equation}
\label{eq:gff}
g_{\varphi\varphi} 
\sim 
-E_{*}\sin^{2}{\theta}\left[1-\left(\frac{u_{*}}{E_{*}}\right)^{2}r^{4}\sin^{2}{\theta}
\right]\, .  
\end{equation}

\begin{figure}[h]
  \centering
  \includegraphics[height=6cm]{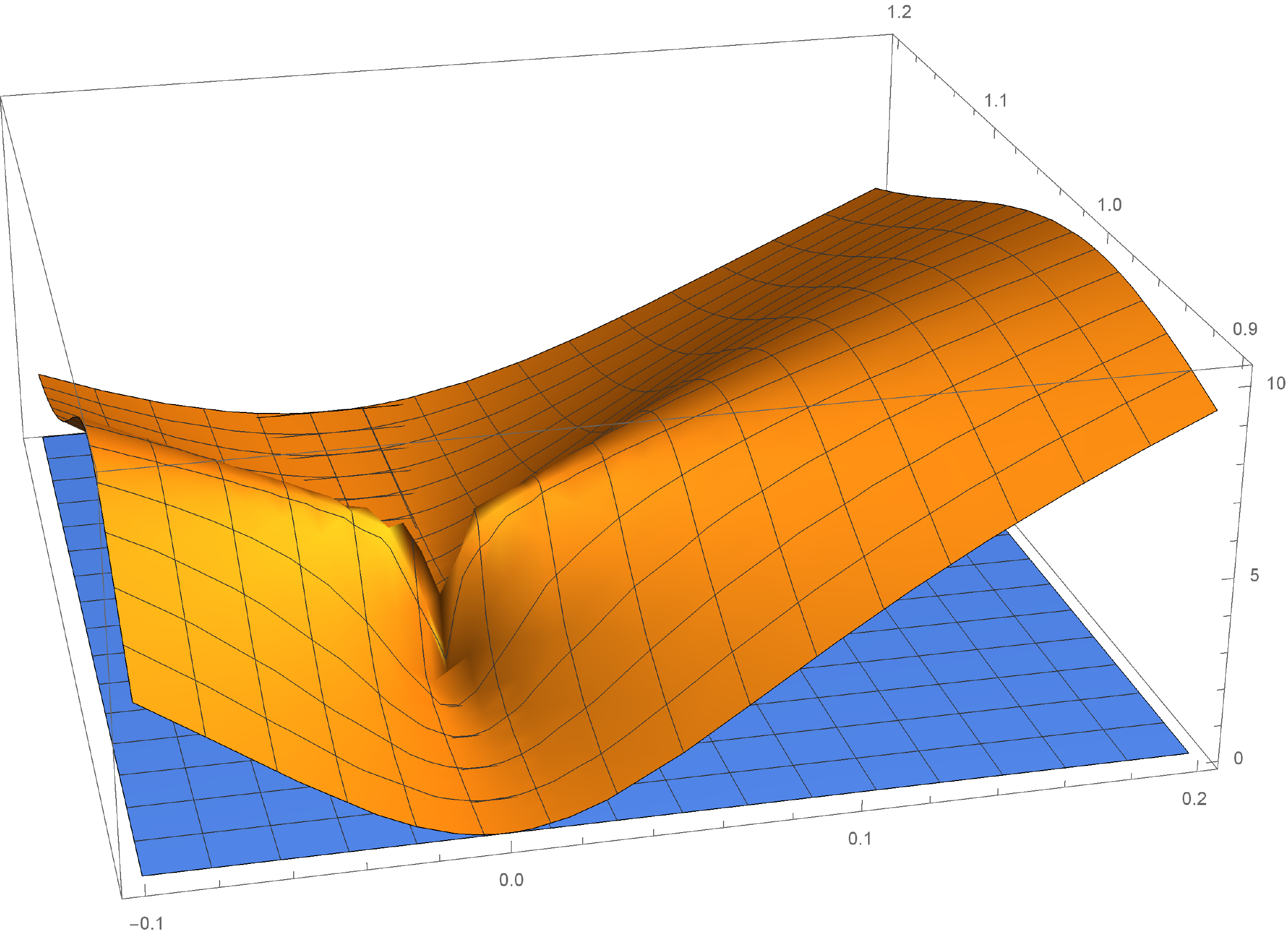}
  \caption{\small The two terms of the r.h.s.~of Eq.~(\ref{eq:gff}) are
    represented as two different surfaces for the case of two centers with
    $\vec{x}_{1}=(0,0,0),\vec{x}_{2}=(0,0,1), h=1,p^{1}=2,p^{2}=3,\lambda =1,
    (s_{\alpha})=(3,2),g=1,(\eta_{\alpha})=(1,2)$. The blue surface always lays below the yellow surface, whence $g_{\varphi\varphi}$ remains finite and positive.}
  \label{fig:graficaprometida}
\end{figure}

In that expression the second term is always smaller than the first term in this limit. Beyond this analysis, we have
explored numerically the value of $g_{\varphi\varphi}$ for several, simple,
multicenter configurations and have found that it can vanish (for instance,in
a 2-center example, all along the the axis that contains both centers) but it
never changes sign. See Fig.~\ref{fig:graficaprometida} for a simple two-center example.

\subsection{Solutions of the $\mathrm{ST}[2,6]$ model}
\label{sec-st26}

\subsubsection{The model}
\label{sec-st25d=4}

The ST$[2,6]$ model is the cubic model with prepotential

\begin{equation}
\label{eq:prepotential}
\mathcal{F}  
= 
-\tfrac{1}{3!}
\frac{d_{ijk}\mathcal{X}^{i}\mathcal{X}^{j}\mathcal{X}^{k}}{\mathcal{X}^{0}}\, ,
\end{equation}

\noindent
where $i=1,2\cdots,6$ labels the vector multiplets and where the fully
symmetric tensor $d_{ijk}$ has as only non-vanishing components

\begin{equation}
d_{1\alpha\beta}= \eta_{\alpha\beta}\, ,
\,\,\,\,\,
\mbox{where}
\,\,\,\,\,
(\eta_{\alpha\beta}) = \mathrm{diag}(+-\dotsm -)\, ,
\,\,\,\,\,
\mbox{and}
\,\,\,\,\,
\alpha,\beta=2,\cdots,6\, .
\end{equation}

\noindent
The 6 complex scalars parametrize the coset space 

\begin{equation}
\frac{\mathrm{SL}(2,\mathbb{R})}{\mathrm{SO}(2)}
\times
\frac{\mathrm{SO}(2,5)}{\mathrm{SO}(2)\times \mathrm{SO}(5)}\, , 
\end{equation}

\noindent
and the group SO$(3)$ acts in the adjoint on the coordinates $\alpha=4,5,6$
that we are going to denote with $A,B,\ldots$ indices. These are the
directions to be gauged. 

In order to construct solutions we only need the Hesse potential of this
theory, which is given by 

\begin{equation}
\label{eq:HessepotentialST25}
W(\mathcal{I})
=
2\sqrt{
(\mathcal{I}^{\alpha}\mathcal{I}^{\beta}\eta_{\alpha\beta}
+2\mathcal{I}^{0}\mathcal{I}_{1})
(\mathcal{I}_{\alpha}\mathcal{I}_{\beta}\eta^{\alpha\beta}
-2\mathcal{I}^{1}\mathcal{I}_{0})
-(\mathcal{I}^{0}\mathcal{I}_{0}-\mathcal{I}^{1}\mathcal{I}_{1}
+\mathcal{I}^{\alpha}\mathcal{I}_{\alpha})^{2}
}\, .
\end{equation}

We could have gauged any three of the directions
$3,4,5,6$, and, therefore, the ungauged one could have been truncated from our
model.  However, as shown in Ref.~\cite{Cano:2016rls}, it is necessary to have
one additional Abelian vector field to be able to uplift the solution to 6
dimensions and then to Heterotic supergravity
\cite{Cano:2017qrq}. Furthermore, with the extra Abelian vector multiplet, the
model can be seen as the STU model coupled to an SU$(2)$ triplet. This can be
made manifest by combining the Abelian directions $2$ and $3$ as follows

\begin{equation}
\mathcal{I}^{\pm}\equiv \mathcal{I}^{2}\pm \mathcal{I}^{3}\, ,
\hspace{1cm}
\mathcal{I}_{\pm}\equiv \mathcal{I}_{2}\pm \mathcal{I}_{3}\, ,  
\end{equation}

\noindent
so that 

\begin{equation}
\eta_{\alpha\beta}\mathcal{I}^{\alpha}\mathcal{I}^{\beta}
=
\mathcal{I}^{+}\mathcal{I}^{-}-\mathcal{I}^{A}\mathcal{I}^{A}\, ,
\hspace{1cm}
\mathcal{I}^{\alpha}\mathcal{I}_{\alpha}
=
\tfrac{1}{2}
\mathcal{I}^{+}\mathcal{I}_{+}
+\tfrac{1}{2}
\mathcal{I}^{-}\mathcal{I}_{-}
+\mathcal{I}^{A}\mathcal{I}_{A}\, .  
\end{equation}

The S, T and U vector fields correspond to the directions $1$, $+$ and $-$,
and the pure STU model is recovered by eliminating all objects with SU$(2)$
indices $A,B,\ldots$

The K\"ahler potential of this model is given by 

\begin{equation}
\label{eq:Kpotential}
e^{-\mathcal{K}}
= 
4\Im\mathfrak{m} Z^{1} \,
\eta_{\alpha\beta}\Im\mathfrak{m} Z^{\alpha}  \Im\mathfrak{m} Z^{\beta}\, ,  
\end{equation}

\noindent
whose positivity leads to a constraint on the possible values of the imaginary parts of
the scalar fields, a constraint that we will use later.

More details on this theory and, in particular, on its relation with the
toroidal compactification of the Heterotic string can be found in
Refs.~\cite{Bueno:2014mea,Meessen:2015enl,Cano:2017sqy,Cano:2017qrq}.

\subsubsection{The solutions}

For the sake of simplicity, we are going to consider solutions with
non-vanishing functions $\mathcal{I}^{0},\mathcal{I}^{A},\mathcal{I}_{1},
\mathcal{I}_{+},\mathcal{I}_{-},\mathcal{I}_{A}$ only.\footnote{These
  solutions, with $\mathcal{I}^{0}\neq 0$ can be uplifted to timelike
  supersymmetric solutions of $\mathcal{N}=1,d=5$ SEYM.} It is convenient to
redefine these functions,

\begin{equation}
\mathcal{I}^{0,1,A} = -\sqrt{2}\Phi^{0,1,A}\, ,
\hspace{.5cm}
\mathcal{I}_{+,-,A} = \tfrac{1}{\sqrt{2}}\mathcal{J}_{+,-,A}\, ,  
\hspace{.5cm}
\mathcal{I}_{1} = -\tfrac{1}{\sqrt{2}}\mathcal{J}_{1}\, ,  
\end{equation}

\noindent
bringing the metric function $e^{-2U}$, the 1-form $\omega$ and the scalar
fields to the form

\begin{eqnarray}
\label{eq:metricold}
e^{-2U}
& = & 
2
\sqrt{
\Phi^{0}\mathcal{J}_{1}\mathcal{J}_{+}\mathcal{J}_{-}
-\mathcal{J}_{+}\mathcal{J}_{-}\vec{\Phi}^{2}
-\Phi^{0}\mathcal{J}_{1}\vec{\mathcal{J}}^{2}
+|\vec{\Phi}\times \vec{\mathcal{J}}|^{2}
}
\\
& & \nonumber \\
\vec{\omega}
& = &
2\, \vec{\Phi}\times \vec{\mathcal{J}}\, ,
\\
& & \nonumber \\
Z^{1}
& \equiv &
\tau 
=
\frac{2(\mathcal{J}_{+}\mathcal{J}_{-}
-\vec{\mathcal{J}}^{2})}{4 \vec{\Phi}\cdot \vec{\mathcal{J}} -i e^{-2U}}\, ,
\\
& & \nonumber \\
Z^{\pm}
& = &
\frac{-2(\mathcal{J}_{\mp}/ \Phi^{0})(\mathcal{J}_{1}\Phi^{0} - \vec{\Phi}^{2})}{4 
\vec{\Phi}\cdot \vec{\mathcal{J}} -i e^{-2U}}\, ,
\\
& & \nonumber \\
\vec{Z}
& = &
\frac{2(\vec{\mathcal{J}}/ \Phi^{0})(\mathcal{J}_{1}\Phi^{0} - \vec{\Phi}^{2})
+4\vec{\Phi}/\Phi^{0} (\vec{\Phi}\cdot\vec{\mathcal{J}})
-i\vec{\Phi}/\Phi^{0}\, e^{-2U})}{4 
\vec{\Phi}\cdot \vec{\mathcal{J}} -i e^{-2U}}\, ,
\end{eqnarray}

\noindent
while the vector fields are given by

\begin{eqnarray}
A^{0}
& = &
-4e^{4U}\Phi^{0} (\vec{\Phi}\cdot\vec{\mathcal{J}})(dt+\omega)\, ,
\\
& & \nonumber \\  
A^{1}
& = &
-2e^{4U}\Phi^{0}(\mathcal{J}_{+}\mathcal{J}_{-}-\vec{\mathcal{J}}^{2})(dt+\omega)\, ,
\\
& & \nonumber \\  
A^{\pm}
& = &
2e^{4U}\mathcal{J}_{\mp} 
(\mathcal{J}_{1}\Phi^{0} -\vec{\Phi}^{2})(dt+\omega)\, ,
\\
& & \nonumber \\  
\vec{A}
& = &
2 e^{4U}
\left\{\vec{\mathcal{J}} (\mathcal{J}_{1}\Phi^{0} -\vec{\Phi}^{2})
+4 \vec{\Phi} (\vec{\Phi}\cdot\vec{\mathcal{J}})
\right\}(dt+\omega)
+
\vec{\breve{A}}\, .
\end{eqnarray}

The explicit magnetic part of the SU$(2)$ vector field, $\vec{\breve{A}}$, is
determined by $\vec{\Phi},\vec{\mathcal{J}}$, which we will choose as in the
$\overline{\mathbb{CP}}^{3}$ model Eqs.~(\ref{eq:14}) and (\ref{eq:PhiJ}). We rewrite them here for convenience:

\begin{equation}
\vec{\Phi}
= 
-\frac{1}{gP}\vec{\nabla}P\, ,
\hspace{.5cm}
\vec{\mathcal{J}}
=
\frac{2}{gP}\vec{\nabla}Q\, ,
\,\,\,\,\,
\Rightarrow
\,\,\,\,\,
\breve{A}^{A}{}_{\underline{r}}
=
-\varepsilon^{A}{}_{rs}\frac{1}{gP}\partial_{\underline{s}}P\, ,  
\end{equation}

\noindent
where

\begin{equation}
P 
=
\lambda+\sum_{\alpha=1}^{N}\frac{s_{\alpha}}{r_{\alpha}}\, ,  
\hspace{1cm}
Q
=
-\sum_{\alpha=1}^{N}\frac{\eta_{\alpha}s_{\alpha}/2}{r_{\alpha}}\, .  
\end{equation}

The Abelian functions
$\Phi^{0},\mathcal{J}_{1},\mathcal{J}_{+},\mathcal{J}_{-}$ will be given by

\begin{equation}
\label{eq:defs}
\Phi^{0} 
=
h^{0} + \sum_{\alpha=1}^{N}\frac{p^{0}_{\alpha}}{r_{\alpha}}
\hspace{1cm}  
\mathcal{J}_{1,\pm} 
=
h_{1,\pm} + \sum_{\alpha=1}^{N}\frac{q_{1,\pm\, \alpha}}{r_{\alpha}}\, .
\end{equation}

The above form of the metric function \eqref{eq:metricold} has the interesting feature that the
1-form $\omega$ appears in it (the last term). If we switch off all the
functions but $\vec{\Phi}$ and $\vec{\mathcal{J}}$, $e^{-2U}=|\omega|$
and we get a metric which is completely determined by $\omega$, but which is
not asymptotically flat neither free of singularities since $|\omega|$ can vanish.

We are going to work with the following alternative form of the metric function

\begin{equation}
e^{-2U}
=
2
\sqrt{
(\mathcal{J}_{1}\Phi^{0} -\vec{\Phi}^{2})
(\mathcal{J}_{+}\mathcal{J}_{-}-\vec{\mathcal{J}}^{2})
-(\vec{\Phi}\cdot\vec{\mathcal{J}})^{2}
}\, .  
\end{equation}

If we plug into Eq.~(\ref{eq:Kpotential}) the values of the scalars, we find
the condition

\begin{equation}
\label{eq:1stconstraint}
\mathcal{J}_{+}\mathcal{J}_{-}-\vec{\mathcal{J}}^{2} > 0\, ,  
\,\,\,\,\,
\Rightarrow
\,\,\,\,\,
\Im\mathfrak{m}\tau >0\, ,
\end{equation}

\noindent
and, using this condition in the above form of the metric function we find a
second regularity condition

\begin{equation}
\label{eq:2ndconstraint}
\mathcal{J}_{1}\Phi^{0} -\vec{\Phi}^{2} > 0\, .  
\end{equation}

These conditions are necessary but not sufficient to ensure the regularity of
the solution, which also requires 

\begin{equation}
\label{eq:3rdconstraint}
(\mathcal{J}_{1}\Phi^{0} -\vec{\Phi}^{2})
(\mathcal{J}_{+}\mathcal{J}_{-}-\vec{\mathcal{J}}^{2})
-(\vec{\Phi}\cdot\vec{\mathcal{J}})^{2} >0\, .
\end{equation}

\subsubsection{Spherically-symmetric and dumbbell solutions}

Again, we start by studying solutions with a single center that we
conveniently place at $\vec{x}=0$, suppressing all indices
$\alpha,\beta,\ldots$ Since $\vec{\Phi}\propto \vec{\mathcal{J}}$ the 1-form
$\omega$ vanishes and the solutions are necessarily static. 

Imposing the standard normalization of the metric at spatial infinity and
studying the asymptotic behavior of the scalar fields we identify the
integration constants $h^{0},h_{1},h_{+},h_{-}$ in Eq.~(\ref{eq:defs}) as

\begin{eqnarray}
h^{0} 
& = & 
\frac{1}{\sqrt{2 \Im\mathfrak{m} \tau_{\infty}
\Im\mathfrak{m} Z_{\infty}^{+}\Im\mathfrak{m} Z_{\infty}^{-}}}\, ,
\hspace{1cm}
h_{1}  
= 
\frac{\Im\mathfrak{m} Z_{\infty}^{+}\Im\mathfrak{m} Z_{\infty}^{-}}{\sqrt{2 
\Im\mathfrak{m} \tau_{\infty}
\Im\mathfrak{m} Z_{\infty}^{+}\Im\mathfrak{m} Z_{\infty}^{-}}}
\, ,
\nonumber \\
\label{eq:hconstants}
& & \\
h_{\pm}  
& = & 
-
\frac{\sqrt{2 \Im\mathfrak{m} \tau_{\infty}
\Im\mathfrak{m} Z_{\infty}^{+}\Im\mathfrak{m} Z_{\infty}^{-}}}{2 
\Im\mathfrak{m} Z_{\infty}^{\mp}}\, ,
\nonumber 
\end{eqnarray}

\noindent
and we will take $q_{1},p^{0}>0$ and $q_{+}q_{-}>0$ with $\mathrm{sign}
(q_{\pm}) = \mathrm{sign}(h_{\pm}) = -\mathrm{sign} (\Im\mathfrak{m}Z^{\mp})$.

Let us consider the first regularity condition
Eq.~(\ref{eq:1stconstraint}). Expanding the functions in the left-hand side we
find

\begin{equation}
\mathcal{J}_{+}\mathcal{J}_{-}-\vec{\mathcal{J}}^{2}
=
h_{+}h_{-}
+
\frac{2A}{r}
+\left[\Sigma + \frac{\eta^{2}}{g^{2}} R(r)\right]
\frac{1}{r^{2}}
\, ,
\end{equation}

\noindent
where, given the values of the $h$ constants,  

\begin{eqnarray}
h_{+}h_{-} 
& = &
\tfrac{1}{2}\Im\mathfrak{m}\tau >0\, ,
\nonumber \\
& & \\
2A 
& = & 
h_{+}q_{-}+h_{-}q_{+}
=
\sqrt{2 \Im\mathfrak{m} \tau_{\infty}
\Im\mathfrak{m} Z_{\infty}^{+}\Im\mathfrak{m} Z_{\infty}^{-}}
\left(
\frac{|q_{+}|}{2|\Im\mathfrak{m} Z_{\infty}^{-}|}
+
\frac{|q_{-}|}{2|\Im\mathfrak{m} Z_{\infty}^{+}|}
\right)
>0\, ,
\nonumber  
\end{eqnarray}

\noindent
$R(r)$ is the non-negative function given in Eq.~(\ref{eq:Rfunction})
so the combination $\Sigma$ must be positive

\begin{equation}
\Sigma \equiv q_{+}q_{-} - \frac{\eta^{2}}{g^{2}}>0\, .  
\end{equation}

Doing the same with the second regularity condition
Eq.~(\ref{eq:2ndconstraint}) we get

\begin{equation}
\mathcal{J}_{1}\Phi^{0} -\vec{\Phi}^{2}
=
h^{0}h_{1}
+\frac{2B}{r}
+\left[ 
\Omega +\frac{1}{g^{2}}R(r)
\right] \frac{1}{r^{2}}\, ,
\end{equation}

\noindent
where

\begin{eqnarray}
h^{0}h_{1}
& = &
\frac{1}{2\Im\mathfrak{m} \tau_{\infty}}\, ,
\nonumber \\
& & \\
2B
& = &
h^{0}q_{1}+h_{1}p^{0}
=
\frac{1}{\sqrt{2 \Im\mathfrak{m} \tau_{\infty}
\Im\mathfrak{m} Z_{\infty}^{+}\Im\mathfrak{m} Z_{\infty}^{-}}}
\left[q_{1} + \Im\mathfrak{m} Z_{\infty}^{+}\Im\mathfrak{m} Z_{\infty}^{-}
  p^{0}\right]
>0\, ,  
\nonumber
\end{eqnarray}

\noindent
which will be manifestly positive if the combination

\begin{equation}
  \Omega \equiv p^{0} q_{1} -\frac{1}{g^{2}} >0\, .  
\end{equation}

Finally, let us consider the third regularity condition
Eq.~(\ref{eq:3rdconstraint}). All the terms that originate in the product of
the first two terms are manifestly positive if $\Omega$ and $\Sigma$ are
positive. The only negative terms come from the last term and are of
$\mathcal{O}(r^{-4})$

\begin{equation}
-(\vec{\Phi}\cdot\vec{\mathcal{J}})^{2}
=
-\frac{\eta^{2}}{g^{2}}[1-R(r)]^{2} \frac{1}{r^{4}}\, .
\end{equation}

We just need to compare them with the positive $\mathcal{O}(r^{-4})$ terms
coming from the first two terms, \textit{i.e.}~we have to consider

\begin{equation}
(\Sigma +\frac{\eta^{2}}{g^{2}}R)(\Omega +\frac{1}{g^{2}}R)  
-\frac{\eta^{2}}{g^{2}}[1-R(r)]^{2}
=
\Omega\Sigma - \frac{\eta^{2}}{g^{4}} 
+\mbox{positive}\,\,\, \mathcal{O}(R)\,\,\, \mbox{terms}.
\end{equation}

Thus, the third regularity condition is fulfilled if we require that

\begin{equation}
E^{2} \equiv \Omega\Sigma - \frac{\eta^{2}}{g^{4}} > 0\, .
\end{equation}

\noindent
Observe that, if this condition is satisfied, the entropy is given by 

\begin{equation}
S = 2\pi E\, .  
\end{equation}
 
\noindent
The conditions that we have imposed on the charges and the asymptotic values
of the scalars automatically ensure the positivity of the mass, which is given
by

\begin{equation}
M 
=  
A+B\, .
\end{equation}

Setting all the $h$ constants to zero, we get a dumbbell solution with metric
function

\begin{equation}
e^{-2U}
=
2
\sqrt{
\left[\Sigma + \frac{\eta^{2}}{g^{2}} R(r)\right]
\left[ \Omega +\frac{1}{g^{2}}R(r)\right]
-\frac{\eta^{2}}{g^{2}}[1-R(r)]^{2}
}
\,\,
\frac{1}{r^{2}}\, ,  
\end{equation}

\noindent
and the square root function interpolates smoothly between $E$ at $r=0$ and
$\sqrt{p^{0}q_{1}q_{+}q_{-}}$ at $r\rightarrow \infty$, which is the value one
would get in the purely Abelian solution. The scalars also interpolate between
a covariant attractor and an Abelian attractor.

Let us now move to the multicenter case.

\subsubsection{Multicenter solutions}

The presence of more centers does not change the asymptotic values of the
scalars and, therefore, the values of the constants $h$ are unchanged and
given by Eqs.~(\ref{eq:hconstants}). We impose on the charges of each center
the same conditions as in the single-center case, that is:

\begin{equation}
\label{eq:cond1}
q_{1\, \alpha},p^{0}_{\alpha}>0\, ,
\,\,\,\,\, 
q_{+\, \alpha}q_{-\, \alpha}>0\, ,
\,\,\,\,\,
\mathrm{sign}(q_{\pm\, \alpha}) 
= 
\mathrm{sign}(h_{\pm}) 
= 
-\mathrm{sign} (\Im\mathfrak{m}Z^{\mp})\, .
\end{equation}

\noindent
Moreover, the four harmonic functions cannot change sign anywhere, as if any of them becomes zero then metric function is imaginary, among other pathologies. Then we can include the conditions

\begin{equation}
\label{eq:cond2}
\mathrm{sign}(q_{1\, \alpha}) 
= 
\mathrm{sign}(h_{1}) 
=
\mathrm{sign}(p^{0}_{\alpha}) 
= 
\mathrm{sign}(h^{0}) \, .
\end{equation}

The first regularity condition Eq.~(\ref{eq:1stconstraint}) can be rewritten
in the form 

\begin{eqnarray}
\mathcal{J}_{+}\mathcal{J}_{-}-\vec{\mathcal{J}}^{2}
& = & 
h_{+}h_{-} +\sum_{\alpha=1}^{N}\frac{2A_{\alpha}}{r_{\alpha}}
+\sum_{\alpha=1}^{N}
\left[\Sigma_{\alpha} 
+\frac{\eta_{\alpha}^{2}}{g^{2}} R_{\alpha}\right]\frac{1}{r^{2}_{\alpha}}
\nonumber \\
& & \nonumber \\
& & 
+\sum_{\alpha<\beta}^{N}
\left[\Sigma_{\alpha\beta}-\Sigma_{\alpha}- \Sigma_{\beta} 
+2\frac{\eta_{\alpha}\eta_{\beta}}{g^{2}}
R_{\alpha\beta}\right]\frac{1}{r_{\alpha}r_{\beta}}>0\, ,
\end{eqnarray}

\noindent
where $R_{\alpha}$ and $R_{\alpha\beta}$ are the functions defined in
Eqs.~(\ref{eq:22}) and (\ref{eq:23}), respectively, and

\begin{eqnarray}
h_{+}h_{-} 
& = &
\tfrac{1}{2}\Im\mathfrak{m}\tau >0\, ,
\\
& & \nonumber \\
2A_{\alpha}  
& \equiv &
h_{+}q_{-\, \alpha}+h_{-}q_{+\, \alpha}>0\, ,
\\
& & \nonumber \\
\Sigma_{\alpha}
& \equiv &
q_{+\, \alpha}q_{-\, \alpha} - \frac{\eta_{\alpha}^{2}}{g^{2}}\, ,
\\
& & \nonumber \\
\Sigma_{\alpha\beta}
& \equiv &
(q_{+\, \alpha}+q_{+\, \beta})
(q_{-\, \alpha}+q_{-\, \beta}) 
- \frac{(\eta_{\alpha}+\eta_{\beta})^{2}}{g^{2}}\, ,
\end{eqnarray}

\noindent
and its positivity is manifest by requiring 

\begin{equation}
\Sigma_{\alpha} > 0\, ,\,\,\, \forall \alpha
\,\,\,\,\,
\mbox{and}
\,\,\,\,\,
\Sigma_{\alpha\beta} > \Sigma_{\alpha}+\Sigma_{\beta}\, ,\,\,\,\, 
\forall \alpha\neq \beta\, .  
\end{equation}

Only the first of these conditions ($\Sigma_{\alpha} > 0\, ,\,\,\, \forall
\alpha$) is independent, though. It implies that $q_{+\, \alpha} >
\frac{\eta_{\alpha}^{2}}{g^{2}q_{-\, \alpha}}$ and, substituting in 

\begin{eqnarray}
\Sigma_{\alpha\beta}-\Sigma_{\alpha}-\Sigma_{\beta} 
& = &  
q_{+\, \alpha}q_{-\, \beta}+ q_{+\, \beta}q_{-\, \alpha} 
-2\frac{\eta_{\alpha}\eta_{\beta}}{g^{2}}
\nonumber \\
& & \nonumber \\
& > &
\frac{\eta^{2}_{\alpha}}{g^{2}q_{-\, \alpha}}q_{-\, \beta}
+\frac{\eta^{2}_{\beta}}{g^{2}q_{-\, \beta}}q_{-\, \alpha} 
-2\frac{\eta_{\alpha}\eta_{\beta}}{g^{2}}
\nonumber \\
& & \nonumber \\
& = & 
\frac{(\eta_{\alpha}q_{-\, \beta}-\eta_{\beta}q_{-\, \alpha})^{2}}{g^{2}q_{-\, \alpha}q_{-\, \beta}} 
\nonumber \\
& & \nonumber \\
& \geq  &
0\, . 
\end{eqnarray}

In a similar way, we rewrite the second condition Eq.~(\ref{eq:1stconstraint})
in the form

\begin{eqnarray}
\mathcal{J}_{1}\Phi^{0} -\vec{\Phi}^{2}
& = &
h^{0}h_{1}
+\sum_{\alpha=1}^{N}\frac{2B_{\alpha}}{r_{\alpha}}
+\sum_{\alpha=1}^{N}\left[ 
\Omega_{\alpha} +\frac{1}{g^{2}}R_{\alpha}
\right] \frac{1}{r_{\alpha}^{2}}
\nonumber \\
& & \nonumber \\
& & 
+\sum_{\alpha<\beta}^{N}\left[ 
\Omega_{\alpha\beta}-\Omega_{\alpha}-\Omega_{\beta} +\frac{2}{g^{2}}R_{\alpha\beta}
\right] \frac{1}{r_{\alpha}r_{\beta}}>0\, ,
\end{eqnarray}

\noindent
where now

\begin{eqnarray}
h^{0}h_{1} 
& = &
\frac{1}{2\Im\mathfrak{m} \tau_{\infty}}>0\, ,
\\
& & \nonumber \\
2B_{\alpha}  
& \equiv &
h^{0}q_{1\, \alpha}+h_{1}p_{\alpha}^{0}>0\, ,
\\
& & \nonumber \\
\Omega_{\alpha}
& \equiv &
p^{0}_{\alpha}q_{1\, \alpha} - \frac{1}{g^{2}}\, ,
\\
& & \nonumber \\
\Omega_{\alpha\beta}
& \equiv &
(p^{0}_{\alpha}+p^{0}_{\beta})
(q_{1\, \alpha}+q_{1\, \beta}) 
- \frac{(1+1)^{2}}{g^{2}}\, .
\end{eqnarray}

\noindent
The positivity bound is obviously satisfied by requiring 

\begin{equation}
\Omega_{\alpha}>0\, ,\,\,\, \forall \alpha
\,\,\,\,\,
\mbox{and}
\,\,\,\,\,
\Omega_{\alpha\beta}>\Omega_{\alpha}+\Omega_{\beta}\, ,\,\,\,\, \forall \alpha\neq \beta\, ,  
\end{equation}

\noindent
and one can show, as before, that the first condition implies the second.

Finally, let us study the third condition Eq.~(\ref{eq:3rdconstraint}). Again,
all the terms that come from the first two factors (corresponding to the first
two conditions) are positive if the conditions that we have derived above are
met. The negative contributions come from\footnote{Observe that this means that, when
  $\eta_{\alpha}=0\, ,\,\,\, \forall \alpha$, or, equivalently, when
  $\vec{\mathcal{J}}=0$ the multicenter solution with non-Abelian magnetic
  monopoles is completely regular. It is only the dyonic case that needs to be
  investigated more carefully.}

\begin{equation}
-(\vec{\Phi}\cdot\vec{\mathcal{J}})^{2}
=
-\frac{1}{g^{4}}\sum_{\alpha, \beta, \gamma, \delta=1}^{N}
\eta_{\alpha}\eta_{\gamma}(1-R_{\alpha\beta})(1-R_{\gamma\delta})
\frac{1}{r_{\alpha}r_{\beta}r_{\gamma}r_{\delta}}\, ,  
\end{equation}

\noindent
and they have to be compared with other (positive) terms of the same order, 
$\mathcal{O}(r^{-4})$ and with the same structure. Let us first consider terms
of the form $r^{-4}_{\alpha}$, which are dominant in the $\alpha^{th}$ near-horizon region,

\begin{equation}
\sum_{\alpha=1}^{N}
\left\{
\left[\Sigma_{\alpha} +\frac{\eta^{2}_{\alpha}}{g^{2}}R_{\alpha}\right]  
\left[\Omega_{\alpha} +\frac{1}{g^{2}}R_{\alpha}\right]
-\frac{\eta^{2}_{\alpha}}{g^{4}}(1-R_{\alpha})^{2}
\right\}
\frac{1}{r_{\alpha}^{4}}\, .
\end{equation}

The positivity of these terms is guaranteed by the positivity of
$\Sigma_{\alpha}$ and $\Omega_{\alpha}$, which we have required before, and
the reality of the entropy of each black hole:

\begin{equation}
\label{eq:cond3}
S_{\alpha}\ =\ 2\pi E_{\alpha} \hspace{1cm}\mbox{with}\hspace{0.5cm}
E^{2}_{\alpha}
\equiv 
\Sigma_{\alpha}\Omega_{\alpha}-\frac{\eta^{2}_{\alpha}}{g^{4}}
>0\, .
\end{equation}

This implies that the metric function is well-defined in the neighbourhood of each black hole, provided the corresponding entropy is real. On the other hand, asymptotic flatness and the sign conditions on the parameters described above, which in turn imply positivity of the ``masses'', guarantee that the metric is also regular far away from any center. However, contrary to our experience with the $\overline{\mathbb{CP}}^3$ model, we have not been able to find a general analytical proof of the regularity of the metric due to the complexity of the $ST[2,6]$ model, as we will shortly see. Nevertheless, we expect most multicenter solutions asymptotically flat and with well-defined individual entropies to be regular everywhere, as the non-Abelian terms generally decay faster with distance than the Abelian harmonic functions.

 Because of their simplicity, let us consider the terms of the form 
$r^{-2}_{\alpha}r^{-2}_{\beta}$:

\begin{eqnarray}
\sum_{\alpha<\beta}^{N}
\left\{
2\left[\Sigma_{(\alpha} +\frac{\eta^{2}_{\alpha}}{g^{2}}R_{(\alpha}\right]  
\left[\Omega_{\beta)} +\frac{1}{g^{2}}R_{\beta)}\right]
\right.
& & \nonumber \\
& & \nonumber \\
+
\left[\Sigma_{\alpha\beta}-\Sigma_{\alpha}- \Sigma_{\beta} 
+2\frac{\eta_{\alpha}\eta_{\beta}}{g^{2}}
R_{\alpha\beta}\right]
\left[ 
\Omega_{\alpha\beta}-\Omega_{\alpha}-\Omega_{\beta} +\frac{2}{g^{2}}R_{\alpha\beta}
\right]
& & \nonumber \\
& & \nonumber \\
\left.
- 
\left[
\frac{2\eta_{\alpha}\eta_{\beta}}{g^{4}}(1-R_{\alpha})(1-R_{\beta})
+\frac{(\eta_{\alpha}+\eta_{\beta})^{2}}{g^{4}}(1-R_{\alpha\beta})^{2}
\right]
\right\}\frac{1}{r_{\alpha}^{2}r_{\beta}^{2}}\, .
& &   
\end{eqnarray}

The coefficient of $r^{-2}_{\alpha}r^{-2}_{\beta}$ has constant terms and other
terms which are linear and quadratic in $R_{\alpha}$ and $R_{\alpha\beta}$.
The linear ones are manifestly positive. The quadratic terms add up to 

\begin{eqnarray}
\frac{(\eta_{\alpha}-\eta_{\beta})^{2}}{g^{4}}
(R_{\alpha}R_{\beta}-R_{\alpha\beta})
\hspace{-5cm}
& & \nonumber \\
& & \\
& = & 
\frac{(\eta_{\alpha}-\eta_{\beta})^{2}}{g^{4}}
\frac{\left[1 - (1+K_{\alpha})^{2}-(1+K_{\beta})^{2}
+2\vec{n}_{\alpha}\cdot\vec{n}_{\beta}(1+K_{\alpha})(1+K_{\beta}) 
-(\vec{n}_{\alpha}\cdot\vec{n}_{\beta})^{2}
\right]}{(1+K_{\alpha})^{2}(1+K_{\beta})^{2}}
\, ,
\nonumber  
\end{eqnarray}

\noindent
which is clearly negative when
$\vec{n}_{\alpha}\cdot\vec{n}_{\beta}=0$. However, it is bounded from above and
below and its negative contribution can still be cancelled by the other terms.

The constant terms are 

\begin{equation}
\Delta_{\alpha\beta} 
+
(\Sigma_{\alpha\beta} -\Sigma_{\alpha} -\Sigma_{\beta})
(\Omega_{\alpha\beta} -\Omega_{\alpha} -\Omega_{\beta})
-
\frac{(\eta_{\alpha}+\eta_{\beta})^{2}}{g^{4}}\, ,
\end{equation}

\noindent
where we have defined

\begin{equation}
\Delta_{\alpha\beta} \equiv   2\Sigma_{(\alpha}\Omega_{\beta)}
-2\frac{\eta_{\alpha}\eta_{\beta}}{g^{2}}\, .
\end{equation}

$\Delta_{\alpha\beta}$ is positive under the assumptions we have made,
because, for instance

\begin{equation}
\Delta_{\alpha\beta}
\geq 
\frac{(\eta_{\beta}\Omega_{\alpha}-\eta_{\alpha}\Omega_{\beta})^{2}}{g^{2}\Omega_{\alpha}\Omega_{\beta}}\, .  
\end{equation}

\noindent
The second term is also positive, but the third is negative. Based on our
previous experience with the $\overline{\mathbb{CP}}^{3}$ model, we can try to
relate this coefficient to the superadditivity of the entropy, rewriting it as
follows:

\begin{eqnarray}
& & 
E^{2}_{\alpha\beta}
-(E_{\alpha}+E_{\beta})^{2}
+2E_{\alpha}E_{\beta}
\nonumber \\
& &  \\
& & 
-
(\Sigma_{\alpha} +\Sigma_{\beta})
(\Omega_{\alpha\beta} -\Omega_{\alpha} -\Omega_{\beta})
-
(\Sigma_{\alpha\beta} -\Sigma_{\alpha} -\Sigma_{\beta})
(\Omega_{\alpha} +\Omega_{\beta})
-
\frac{(\eta_{\alpha}+\eta_{\beta})^{2}}{g^{4}}\, .  
\nonumber
\end{eqnarray}

This expression is not very enlightening as there is no simple way to show
that the would-be positive terms in the first line are actually larger than
the negative ones in the second.\footnote{The superadditivity condition
  $E_{\alpha\beta}\geq E_{\alpha}+E_{\beta}$ or $E^{2}_{\alpha\beta}
  -(E_{\alpha}+E_{\beta})^{2}\geq 0$ does not seem to lead to any identity
  that can be used directly in the terms at hands. From the conditions
  $\Sigma_{\alpha\beta}\geq \Sigma_{\alpha}+\Sigma_{\beta}$ and
  $\Omega_{\alpha\beta}\geq \Omega_{\alpha}+\Omega_{\beta}$ we find 

\begin{eqnarray}
E^{2}_{\alpha\beta}
& = &
\Sigma_{\alpha\beta}\Omega_{\alpha\beta} 
-\frac{(\eta_{\alpha}+\eta_{\beta})^{2}}{g^{4}}
\nonumber \\
& & \nonumber \\
& \geq &
(\Sigma_{\alpha}+\Sigma_{\beta})(\Omega_{\alpha}+\Omega_{\beta})
-\frac{(\eta_{\alpha}+\eta_{\beta})^{2}}{g^{4}}
\nonumber \\
& & \nonumber \\
& = &
E^{2}_{\alpha}+E^{2}_{\beta} +\Delta_{\alpha\beta}\, .
\end{eqnarray}

\noindent
Adding and substracting $2E_{\alpha}E_{\beta}$ we arrive to 

\begin{equation}
E^{2}_{\alpha\beta}
-(E_{\alpha} + E_{\beta})^{2}
\geq 
\Delta_{\alpha\beta}-2E_{\alpha}E_{\beta}\, ,
\end{equation}

\noindent
which cannot be used for our purposes.

}

Summarizing, not all the terms that appear below the square root sign in the
metric function are positive definite and we have not been able to determine a set of
conditions ensuring the positivity of the whole expression and the regularity
of the metric function, which still might possible, in accordance with our experience with the
$\overline{\mathbb{CP}}^{3}$ model. 

To conclude this subsection we are going to give an explicit example of a completely regular two-center dyonic
solution of this model. Our choice of charges for the two centers is given in
Table~\ref{tab:2-centerexample}. In this case, the coefficient of the
$r_{1}^{-2}r_{2}^{-2}$ term is the only one which is not manifestly positive
and is given by

\begin{equation}
\frac{17}{2} +\frac{11}{4}R_{1} +\frac{13}{4}R_{2} +\frac{27}{2}R_{12} 
+\frac{1}{4}R_{1}R_{2}-\frac{1}{4}R_{12}^{2}\, .  
\end{equation}

\noindent
However, since $R_{12}^{2}\leq 1$, this term is positive everywhere.

Observe that the 1-form $\omega$ has exactly the same form as in the
$\overline{\mathbb{CP}}^{3}$ model case and, as the analysis made in that case
showed, it will have no effect on the regularity of the metric.

\begin{table}
\centering
\begin{tabular}{||c||c|c|c|c|c|c|c|c||}
\hline\hline  
& $p^{0}$ & $q_{1}$ & $\Omega$ & $q_{+}$ & $q_{-}$ & $\eta$ & $\Sigma$ &  $E^{2}$  \\
\hline\hline  
& & & & & & & & \\
Center 1 & 2 & 1 & 1 & 3 & 1 & 1 & 2 &  1 \\
& & & & & & & & \\
\hline
& & & & & & & & \\
Center 2 & 1 & 2 & 1 & 1 & 1 & 1/2 & 3/4 &  1/2 \\
& & & & & & & & \\
\hline\hline   
\end{tabular}
\caption{Charges and other quantities of the 2-center dyonic solution of the
  ST$[2,6]$ model. We have set the YM coupling constant $g=1$. For this
  solution $\Sigma_{12}=\Omega_{12}=3$, $E^{2}_{12}=27/4$ and
  $E_{12}-E_{1}-E_{2}\sim 0.9 >0$.}
  \label{tab:2-centerexample}
\end{table}

\subsubsection{Thousands of dyonic black holes}
\label{sec-sponge}
While we have not been able to prove the reality of the metric function for completely general configurations, we have argued that most solutions (if not all) described by our construction are well-behaved, provided the ``masses'' and entropies of the individual black holes are positive and real. To add further support to this thesis, we now describe a very general solution composed of $6060$ black holes whose regularity we have checked by numerical analysis. 

\begin{figure}[h]
  \centering
  \includegraphics[height=10cm]{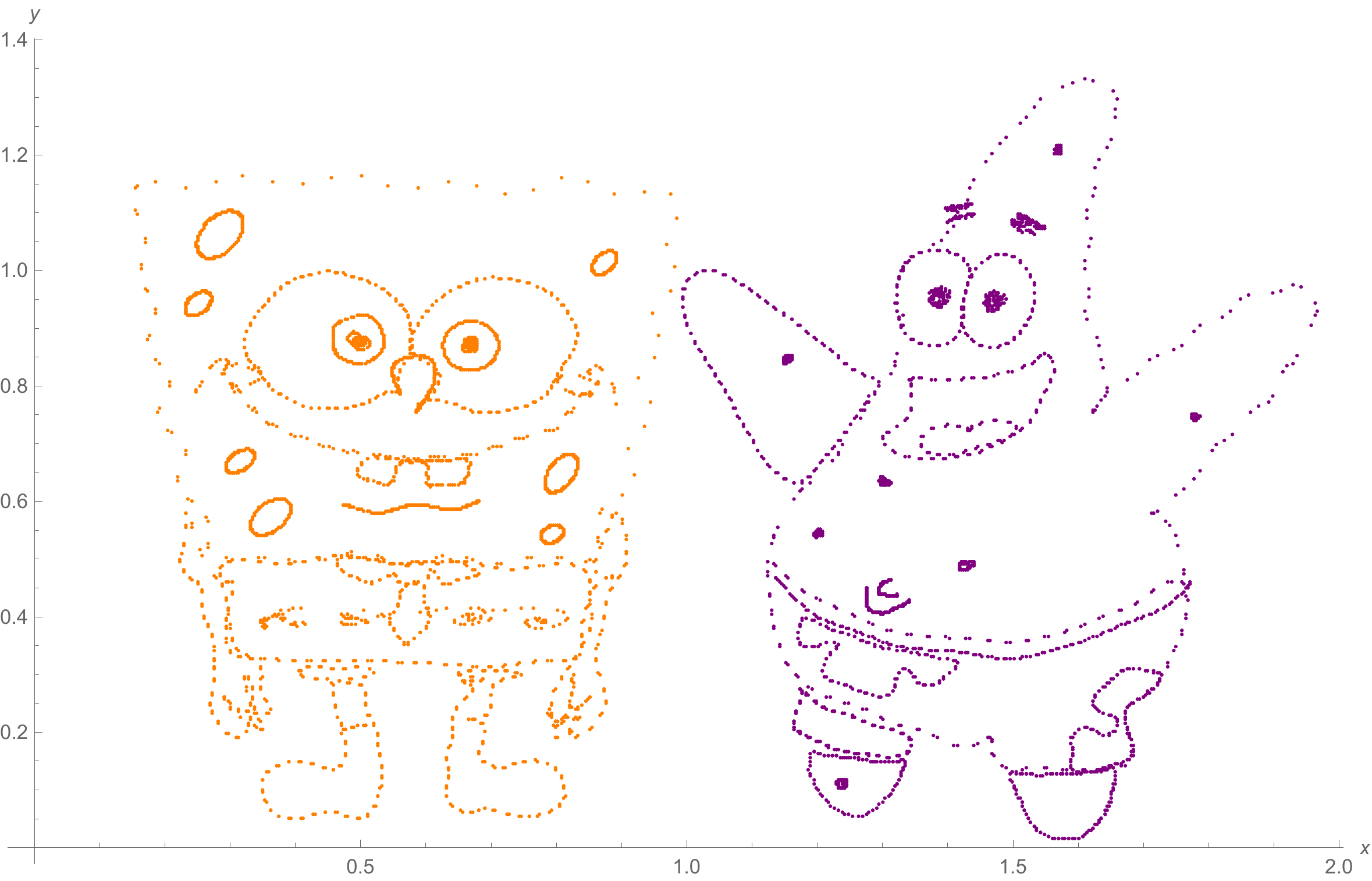}
  \caption{\small Representation of the positions of the black holes, which are contained at the plane $\vec{x}=(x,y,0)$. The first cluster is depicted by purple points, while the second is represented with orange points.}
  \label{fig:sponge}
\end{figure}

The system is composed of two well differentiated clusters. The first cluster describes a set of $1480$ pairs of black holes with the same charges as the two-center system presented at the end of previous section. The second cluster contains $3100$ black holes whose charges have been chosen with a random generator, provided the conditions \eqref{eq:cond1}, \eqref{eq:cond2} and \eqref{eq:cond3} are met. Since the position of each black hole is free, those have been placed as depicted in Figure \ref{fig:sponge} for aesthetic reasons\footnote{Further information about this solution, including a .nb document with the numerical computations, is available upon request by email.}.

\section{Conclusions}
\label{sec-conclusions}

In this paper we have constructed and studied the very first multicenter
black-hole solutions with non-trivial non-Abelian fields corresponding to
\textit{colored} monopoles and dyons. These solutions describe regular black
holes in equilibrium when certain conditions (which we discuss below) are
met. In general, they are stationary, although they have vanishing angular
momentum unless the Abelian fields contribute to it. If these Abelian
contributions are absent, the black holes can be have arbitrary positions.

The main ingredients in the construction of these solutions are

\begin{enumerate}

\item Unbroken supersymmetry, which provides us with a very powerful
  solution-generating technique \cite{Huebscher:2007hj}. The use of this
  technique is only possible if one considers (as we have done here) the
  simplest $\mathcal{N}>1$ supersymmetric generalizations of the
  Einstein-Yang-Mills system. As a reward for considering this generalization
  the solutions are obtained in a completely analytical form. This, in its
  turn, allows for a deeper understating of the solutions.

\item The multi-colored dyon solution of Ref.~\cite{Ramirez:2016tqc}, which
  is the main building block of the physical fields of the 4-dimensional
  spacetime solution. This solution solves the integrability equations
  (\ref{eq:integrability}) everywhere independently of the positions of the
  centers.

\end{enumerate}

Profiting from the analytical form of the metrics obtained, we have tried to
determine general conditions on the charges and moduli guaranteeing
regularity. In the $\mathbb{CP}^{3}$ model with any number of dyonic centers
at arbitrary positions, we have shown that the positivity of each of the
``masses'' and entropies and the superadditivity condition for every pair of
black holes are sufficient to guarantee regularity. Actually, as we have seen
for just two centers, a condition weaker than superadditivity can also be
sufficient. In the ST$[2,6]$ model with only some Abelian vectors active, we
have not been able to prove that similar conditions for an arbitrary
number of dyonic centers are sufficient, although we have explicitly constructed and checked numerically highly non-trivial regular solutions with thousands of black holes. Also, we have shown that very simple conditions are sufficient when there are
only magnetic monopoles at the centers. 

We have also found that, removing the constant part of the harmonic functions
in the spherically symmetric (single-center) solutions one can obtain
solutions that interpolate between two aDS$_{2}\times$S$^{2}$ vacua with
different radii that we have called \textit{dumbbell solutions}. They are the
4-dimensional version of similar 6-dimensional solutions found in
Ref.~\cite{Cano:2016rls} interpolating between two aDS$_{3}\times$S$^{3}$
vacua with different radii, also in a non-Abelian context. The existence of
these solutions in the non-Abelian case\footnote{In the Abelian case, removing the constant part of
  the harmonic functions leads to solutions describing one
  aDS$_{2}\times$S$^{2}$ vacuum in the spherically-symmetric case or
  interpolating between three or more aDS$_{2}\times$S$^{2}$ vacua with
  different radii, but never between just two.} suggests the possible
existence of an Euclidean instanton describing the decay of one vacuum into
the other one. The aDS/CFT interpretation of the corresponding transition (if
found) should provide interesting insights into this correspondence.

As we have discussed in Section~\ref{sec-setup} solutions to the same three
sets of equations (\ref{eq:B})-(\ref{eq:integrability}) can be used to
construct timelike supersymmetric solutions of $\mathcal{N}=1,d=5$ SEYM
theories using different rules to relate the building blocks that occur in
those equations and the physical 5-dimensional fields. Typically, building
blocks that lead to regular 4-dimensional solutions produce singular
5-dimensional solutions and \textit{vice versa}. This means that the
construction of 5-dimensional solutions will have to be studied
independently. Work in this direction is well under way \cite{kn:MOR}.

We have deliberately set aside for future work (already in progress
\cite{kn:CMOR}) the paradoxes created by the strange properties of the
colored dyons which do not seem to contribute to the mass or any other
asymptotic charge (so they behave as non-Abelian hair) but, nevertheless, do
seem to contribute to the entropy. In the 5-dimensional case an analogous
paradox was completely solved in Ref.~\cite{Cano:2017qrq} by the correct,
string theory-inspired, reinterpretation of the Abelian charges and the
identification of a globally regular solution supported by the non-Abelian
field (a BPST instanton) \cite{Cano:2017sqy}. Although we have not yet found
globally regular solutions associated to the 4-dimensional colored dyons, we
expect a similar resolution for this paradox, at least in the case of the
ST$[2,6]$ model, because the string theory embedding of the
$\overline{\mathbb{CP}}^{3}$ model is unknown (or inexistent).\footnote{It
  goes without saying that the numerical character of the solutions of the EYM
  and EYMH models makes them entirely unsuitable for this kind of analysis.}

As mentioned in the introduction, the non-Abelian asymptotically-aDS case is
much harder to deal with in SEYM theories. We are currently working on the
generalization of the methods and solutions used here and we expect to report
on our results soon \cite{kn:COR}.

To conclude, SEYM theories provide new tools to study the interplay between
non-Abelian Yang-Mills and gravitational fields through the construction of a
wealth of new, fully analytical solutions, some of which can be reinterpreted
in the framework of string theory. As we have discussed, there are many
directions to be explored and it is our purpose to follow some of them in the
near future.

\section*{Acknowledgments}

The authors would like to thank C.S.~Shahbazi for interesting
conversations. PFR would like to thank the IPhT for its hospitality and
financial support. This work has been supported in part by the Principado de
Asturias grant GRUPIN14-108, the MINECO/FEDER, UE grant FPA2015-66793-P and
the Centro de Excelencia Severo Ochoa Program grant SEV-2012-0249.  The work
of PFR was further supported by the \textit{Severo Ochoa} pre-doctoral grant
SVP-2013-067903, the EEBB-I-16-11563 grant and the John Templeton Foundation
grant 48222.  TO wishes to thank M.M.~Fern\'andez for her permanent support.

\appendix

\section{$\mathcal{N}=2$, $d=4$ SEYM theories}
\label{app-N2D4SEYM}

$\mathcal{N}=2,d=4$ Super-Einstein-Yang-Mills (SEYM) theories can be seen as
the simplest $\mathcal{N}=2$ supersymmetrization of the Einstein-Yang-Mills
(EYM) theories. They are nothing but theories of $\mathcal{N}=2$, $d=4$
supergravity coupled to $n$ vector multiplets in which a (necessarily
non-Abelian) subgroup of the isometry group of the (Special K\"ahler) scalar
manifold has been gauged using some of the vector fields of the theory as
gauge fields.\footnote{here we are giving a minimal review of these
  theories. More details can be found in
  Refs.~\cite{Hubscher:2008yz,Freedman:2012zz,Ortin:2015hya}; our conventions
  are those of Refs.~\cite{Huebscher:2007hj,Hubscher:2008yz,Ortin:2015hya}.}
The necessary and sufficient conditions for the gauging of a non-Abelian
subgroup of the global symmetry group to be possible are:

\begin{enumerate}
\item It must act on the vector fields in the adjoint representation.
\item It must be a symmetry of the prepotential; see {\em e.g.\/}
  Ref.~\cite{Huebscher:2007hj} for more details.
\end{enumerate}

We will only be concerned with the bosonic sector of the theory, which
consists on the metric $g_{\mu\nu}$, the vector fields
$A^{\Lambda}{}_{\mu}$ ($\Lambda=0,1,\cdots,n$) and the complex scalars
$Z^{i}$ ($i=1,\cdots,n$). The action of the bosonic sector reads

\begin{equation}
\label{eq:N2d4SEYMaction}
  \begin{array}{rcl}
  S[g_{\mu\nu}, A^{\Lambda}{}_{\mu}, Z^{i}] 
  & = & 
{\displaystyle\int} d^{4}x \sqrt{|g|}
  \left[R 
    +2\mathcal{G}_{ij^{*}}\mathfrak{D}_{\mu}Z^{i}\mathfrak{D}^{\mu}Z^{*\, j^{*}}
    +2\Im\mathfrak{m}\mathcal{N}_{\Lambda\Sigma} 
    F^{\Lambda\, \mu\nu}F^{\Sigma}{}_{\mu\nu}
  \right. \\
  & & \\
  & & \left. 
    -2\Re\mathfrak{e}\mathcal{N}_{\Lambda\Sigma}  
    F^{\Lambda\, \mu\nu}\star F^{\Sigma}{}_{\mu\nu}
    -V(Z,Z^{*})
  \right]\, .
\end{array}
\end{equation}

\noindent
In this expression, $\mathcal{G}_{ij^{*}}$ is the K\"ahler
metric, $\mathfrak{D}_{\mu}Z^{i}$ is the gauge-covariant derivative 

\begin{equation}
\mathfrak{D}_{\mu}Z^{i}
= 
\partial_{\mu}Z^{i}+gA^{\Lambda}{}_{\mu}k_{\Lambda}{}^{i}\, ,  
\end{equation}

\noindent
$F^{\Lambda}{}_{\mu\nu}$ is the vector field strength

\begin{equation}
F^{\Lambda}{}_{\mu\nu} 
= 
2\partial_{[\mu}A^{\Lambda}{}_{\nu]} 
+gf_{\Sigma\Gamma}{}^{\Lambda}A^{\Sigma}{}_{\mu}A^{\Gamma}{}_{\nu}\, ,
\end{equation}

\noindent
$\mathcal{N}_{\Lambda\Sigma}$ is the period matrix and, finally, $V(Z,Z^{*})$ is
the scalar potential

\begin{equation}
V(Z,Z^{*}) 
=
-{\textstyle\frac{1}{4}} g^{2}
\Im\mathfrak{m}\mathcal{N}^{\Lambda\Sigma}\mathcal{P}_{\Lambda}\mathcal{P}_{\Sigma}\, .
\end{equation}

Since the imaginary part of the period matrix is negative definite,
the scalar potential is positive semidefinite, which leads to
asymptotically-flat or -De Sitter solutions.

In the above equations, $k_{\Lambda}{}^{i}(Z)$ are the holomorphic Killing
vectors of the isometries that have been gauged\footnote{ The employed
  notation associates a Killing vector to each value of the index $\Lambda$ in
  order to avoid the introduction of yet another class of indices and the
  embedding tensor (See {\em e.g.\/} the reviews \cite{Trigiante:2007ki}); it
  is understood that not all the $k_{\Lambda}$ need to be non-vanishing.  }
and $\mathcal{P}_{\Lambda}(Z,Z^{*})$ the corresponding momentum maps, which
are related to the Killing vectors and to the K\"ahler potential $\mathcal{K}$
by

\begin{eqnarray}
i\mathcal{P}_{\Lambda}
& = &
k_{\Lambda}{}^{i}\partial_{i}\mathcal{K} -\lambda_{\Lambda}\, ,   
\\
& & \nonumber \\
k_{\Lambda\, i^{*}} 
& = &
i\partial_{i^{*}}\mathcal{P}_{\Lambda}\, ,
\end{eqnarray}

\noindent
for some holomorphic functions $\lambda_{\Lambda}(Z)$.  Furthermore,
the holomorphic Killing vectors and the generators $T_{\Lambda}$ of
the gauge group satisfy the Lie algebras

\begin{equation}
[k_{\Lambda},k_{\Sigma}] = -f_{\Lambda\Sigma}{}^{\Gamma}k_{\Gamma}\, ,
\hspace{1cm}
[T_{\Lambda},T_{\Sigma}] = +f_{\Lambda\Sigma}{}^{\Gamma}T_{\Gamma}\, .
\end{equation}

For the gauge group $\mathrm{SU}(2)$, which is the only one we are going to
consider here, we use lowercase indices\footnote{These will be a certain
  subset of those represented by $\Lambda, \Sigma,\ldots$.}  $x,y,z=1,2,3$ and
the structure constants are $f_{xy}{}^{z}=\varepsilon_{xyz}$, so

\begin{equation}
\label{eq:Lieantisu2}
[k_{x},k_{y}] = -\varepsilon_{xyz}k_{z}\, ,
\hspace{1cm}
[T_{x},T_{y}] = +\varepsilon_{xyz}T_{z}\, . %
\end{equation}

The equations of motion of the theory can be written in the following form:

\begin{eqnarray}
\label{eq:Emnn2}
G_{\mu\nu}
+2\mathcal{G}_{ij^{*}}[\mathfrak{D}_{(\mu}Z^{i} \mathfrak{D}_{\nu )}Z^{*\, j^{*}}
-{\textstyle\frac{1}{2}}g_{\mu\nu}
\mathfrak{D}_{\rho}Z^{i}\mathfrak{D}^{\rho}Z^{*\, j^{*}}]
\nonumber \\
& & \nonumber \\
+4 \mathcal{M}_{MN}
\mathcal{F}^{M}{}_{\mu}{}^{\rho}\mathcal{F}^{N}{}_{\nu\rho}
+{\textstyle\frac{1}{2}}g_{\mu\nu}V(Z,Z^{*})
& = & 0,
\\
& & \nonumber \\
\label{eq:Ei2} 
\mathfrak{D}^{2}Z^{i} 
+\partial^{i}G_{\Lambda\, \mu\nu}\star F^{\Lambda\, \mu\nu}
+{\textstyle\frac{1}{2}} \partial^{i}V(Z,Z^{*})
& = & 
0,
\\
& & \nonumber \\
\label{eq:EmL}
\mathfrak{D}_{\nu} \star G_{\Lambda}{}^{\nu\mu}
+{\textstyle\frac{1}{4}}g
\left(
k_{\Lambda\, i^{*}}\mathfrak{D}_{\mu} Z^{*}{}^{i^{*}}+
k^{*}_{\Lambda\, i}\mathfrak{D}_{\mu} Z^{i} 
\right)
& = &
0\, ,
\end{eqnarray}

\noindent
where $G_{\Lambda\, \mu\nu}$ is the dual vector field strength

\begin{equation} 
G_{\Lambda}
\equiv 
\Re\mathfrak{e}\mathcal{N}_{\Lambda\Sigma} F^{\Sigma} 
+
\Im\mathfrak{m}\mathcal{N}_{\Lambda\Sigma}\, \star  F^{\Sigma}\,  ,
\end{equation}

\noindent
$\mathcal{F}^{M}{}_{\mu\nu}$ is the symplectic vector of vector field
strengths

\begin{equation}
\left(\mathcal{F}^{M}\right) 
\equiv 
\left(
\begin{array}{c}
F^{\Lambda} \\ G_{\Lambda} \\ 
\end{array}
\right)\, ,
\end{equation}

\noindent
$\mathcal{M}_{MN}$ is the symmetric $2(n+1)\times 2(n+1)$ matrix
defined  by

\begin{equation}
(\mathcal{M}_{MN})
\equiv 
\left(
  \begin{array}{cc}
\Im\mathfrak{m}\mathcal{N}_{\Lambda\Sigma} 
+R_{\Lambda\Gamma}\Im\mathfrak{m}\mathcal{N}^{-1|\Gamma\Omega}R_{\Omega\Sigma}\,\,\,\,
& 
-R_{\Lambda\Gamma}\Im\mathfrak{m}\mathcal{N}^{-1|\Gamma\Sigma} 
\\
& \\
-\Im\mathfrak{m}\mathcal{N}^{-1|\Lambda\Omega}R_{\Omega\Sigma} 
& 
\Im\mathfrak{m}\mathcal{N}^{-1|\Lambda\Sigma}
\\
\end{array}
\right)\, ,
\end{equation}

\noindent
and 

\begin{equation}
\mathfrak{D}_{\nu} \star G_{\Lambda}{}^{\nu\mu}  
=
\partial_{\nu} \star G_{\Lambda}{}^{\nu\mu}
+
gf_{\Lambda\Sigma}{}^{\Gamma}A^{\Sigma}{}_{\nu}\star G_{\Lambda}{}^{\nu\mu}\, .
\end{equation}

\section{Supersymmetric multi-BH's in pure EM theory}
\label{sec:MinSugr}

Einstein-Maxwell gravity is equivalent to minimal $\mathcal{N}=2,d=4$
supergravity (in fact it could be called the $\overline{\mathbb{CP}}^{0}$
model). The timelike supersymmetric solutions of this supergravity theory are
nothing but the Perj\'es-Israel-Wilson family of solutions
\cite{Perjes:1971gv,Israel:1972vx} which we can, then, study using the
language and methods we use in other models of $\mathcal{N}=2,d=4$
supergravity in the main text, recovering Hartle and Hawking's result
\cite{Hartle:1972ya} that the only regular solutions in this family of
solutions are those of the Majumdar-Papapetrou subfamily
\cite{Majumdar:1947eu,kn:P}. Our starting point will be that one can only use
in the construction of regular solutions harmonic functions with point-like
singularities corresponding to electric or magnetic monopoles, but no higher
multiplets of the electromagnetic field \cite{Bellorin:2006xr}.

The metric function of pure supergravity is given by 

\begin{equation}
\label{eq:60}
e^{-2U}
=
\tfrac{1}{2}\left(\mathcal{I}^{0}\right)^{2}\ 
+\ 2\left(\mathcal{I}_{0}\right)^{2}
=
|\mathcal{H}|^{2}\, ,
\end{equation}

\noindent
where we have defined

\begin{equation}
\mathcal{H}
\equiv  
\tfrac{1}{\sqrt{2}}(\mathcal{I}^{0}+2i\mathcal{I}_{0})\, .
\end{equation}

\noindent
By assumption, the complex function $\mathcal{H}$ has the form 

\begin{equation}
\mathcal{H}
=
h +\sum_{\alpha=1}^{N}\frac{\Gamma_{\alpha}}{r_{\alpha}}\, ,
\,\,\,\,\,
\mbox{where}
\,\,\,\,\,
r_{\alpha} \equiv |\vec{x}-\vec{x}_{\alpha}|\, ,
\end{equation}

\noindent
and the metric function, conveniently normalized at infinity ($h=e^{i\gamma}$)
can be written in the form 

\begin{equation}
\label{eq:62}
e^{-2U} 
= 
1\ +\ \sum_{\alpha}\frac{2M_{\alpha}}{r_{\alpha}}\ 
+\ \sum_{\alpha}\frac{E_{\alpha}}{r_{\alpha}^{2}}\ 
+\ \sum_{\alpha>\beta}(E_{\alpha\beta}-E_{\alpha}-E_{\beta})\frac{1}{r_{\alpha}r_{\beta}}\; ,
\end{equation}

\noindent
where

\begin{equation}
M_{\alpha} 
\equiv 
\Re\mathfrak{e}(e^{i\gamma}\Gamma^{*}_{\alpha})\, ,  
\end{equation}

\noindent
is the mass of the $\alpha^{th}$ black hole,

\begin{equation}
E_{\alpha} 
 \equiv 
 |\Gamma_{\alpha}|^{2}\, , 
\end{equation}

\noindent
is (up to a factor) the entropy of the $\alpha^{th}$ black hole, and 

\begin{equation}
E_{\alpha\beta} 
 \equiv
|\Gamma_{\alpha}+\Gamma_{\beta}|^{2}\, ,
\end{equation}

\noindent
is (up to a factor) the entropy of a black hole with the charges of the
$\alpha^{th}$ and $\beta^{th}$ black holes combined.

It is evident that the metric function will be regular if the masses are
non-negative $M_{\alpha}\geq 0$, the entropies corresponding to centers with
non-vanishing mass are strictly positive $E_{\alpha}>0$ and the entropy of the
combination of two black holes is not smaller than the sum of the entropies of
the individual black holes $E_{\alpha\beta}\geq E_{\alpha}+E_{\beta}$. Given
the expressions for the masses and entropies, it is also evident that the
condition $E_{\alpha}>0$ for $M_{\alpha}>0$ is, actually, redundant.

We also have to examine Eq.~(\ref{eq:integrability}), which, in terms of the
complex function $\mathcal{H}$ takes the form

\begin{equation}
\Im\mathfrak{m}
\left\{
\mathcal{H}\partial_{\underline{r}}\partial_{\underline{r}}\mathcal{H}^{*}
\right\}
=
0\, , 
\end{equation}

\noindent
everywhere. This equation is non-trivial at the locations of the singularities
of the harmonic function $\mathcal{H}$ and leads to the conditions

\begin{equation}
\Im\mathfrak{m}
\left\{
e^{i\gamma}\Gamma^{*}_{\alpha} + \sum_{\beta\neq
  \alpha}\frac{\Gamma^{*}_{\alpha}\Gamma_{\beta}}{r_{\alpha\beta}}
\right\}
=
0\, ,
\,\,\,\,\,
\forall \alpha
\,\,\,\,\,
\mbox{where}
\,\,\,\,\,
r_{\alpha\beta} 
=
|\vec{x}_{\alpha}-\vec{x}_{\beta}|\, .
\end{equation}

Defining the contribution to the total NUT charge of the $\alpha^{th}$ black
hole by

\begin{equation}
N_{\alpha}
\equiv
\Im\mathfrak{m}
(e^{i\gamma}\Gamma^{*}_{\alpha}) \, , 
\end{equation}

\noindent
the above equations can be written in the form

\begin{equation}
\label{eq:bubleMN}
N_{\alpha} 
\left[1
+ 
\sum_{\beta\neq \alpha}\frac{M_{\beta}}{r_{\alpha\beta}}  
\right]
=
M_{\alpha}
\sum_{\beta\neq \alpha}\frac{N_{\beta}}{r_{\alpha\beta}}\, ,  
\end{equation}

\noindent
and the sum over $\alpha$ gives the condition

\begin{equation}
\label{eq:SN=0}
\sum_{\alpha}N_{\alpha}=0\, .  
\end{equation}

\noindent
Furthermore, the condition $E_{\alpha\beta}\geq E_{\alpha}+E_{\beta}$ is
equivalent to

\begin{equation}
\label{eq:MMNN}
M_{\alpha}M_{\beta}+N_{\alpha}N_{\beta} \geq 0\, .  
\end{equation}

\noindent
This condition is also trivially valid for $\alpha=\beta$, which corresponds
to the condition $E_{\alpha}\geq 0$.

For two black holes $N_{2}=-N_{1}$ and the $\alpha=1$ equation takes the form

\begin{equation}
N_{1} \left\{1+ (M_{1}+M_{2})\frac{1}{r_{12}} \right\} =0\, ,  
\end{equation}

\noindent
which, if the masses are positive, as required by the regularity of $e^{-2U}$,
is only be solved by $N_{1}=N_{2}=0$ so the phases of $\Gamma_{1}$ and
$\Gamma_{2}$ are both equal to $e^{i\gamma}$. Then,

\begin{equation}
\mathcal{H}
=
e^{i\gamma}\left( 1 +\frac{|\Gamma_{1}|}{r_{1}}
  +\frac{|\Gamma_{2}|}{r_{2}}\right)\, ,
\end{equation}

\noindent
and the 1-form $\omega$ vanishes identically.

For three black holes, if one of the $N_{\alpha}$ vanishes, we recover the
equations of the two-black-hole case, and the same conclusion. Let us, then,
consider the case in which the three $N_{\alpha}$ are different from
zero. Eqs.~(\ref{eq:bubleMN}) imply that the three masses are also different
form zero.  Due to Eq.~(\ref{eq:SN=0}), two of the $N_{\alpha}$ will have the
same sign and the third will have the opposite sign. With no loss of
generality we can consider $N_{1}>0$ and $N_{2,3}<0$ (the other case differs
only in a global sign). This means that 

\begin{equation}
\sum_{\beta\neq 1} \frac{N_{\beta}}{r_{1\beta}}
=
\frac{N_{2}}{r_{12}} +\frac{N_{3}}{r_{13}}<0\, ,  
\end{equation}

\noindent
and the first of Eqs.~(\ref{eq:bubleMN}) ($\alpha=1$) cannot be satisfied.

The 3 black hole case suggests the way forward for an arbitrary number of
black holes: we can take the sum of all the Eqs.~(\ref{eq:bubleMN}) for which
$N_{\alpha}>0$. Taking into account the cancellations in both sides of the
resulting equation, we get the equation

\begin{equation}
\sum_{\alpha\, |\, N_{\alpha}>0 } N_{\alpha} 
\left[1
+ 
\sum_{\beta | N_{\alpha}<0}\frac{M_{\beta}}{r_{\alpha\beta}}  
\right]
=
\sum_{\alpha | N_{\alpha}>0 }M_{\alpha}
\sum_{\beta | N_{\beta} <0}\frac{N_{\beta}}{r_{\alpha\beta}}\, ,  
\end{equation}

\noindent
whose l.h.s. and r.h.s. are, respectively, positive and negative definite by
assumption.


\end{document}